\documentclass[modern]{aastex63}

\usepackage{afterpage}
\usepackage{hyperref}

\begin{document}

\title{From supernova to supernova remnant:\\ 
comparison of thermonuclear explosion models}

\shorttitle{SN2SNR grid}
\shortauthors{Ferrand et al.}

\correspondingauthor{Gilles Ferrand}
\email{gilles.ferrand@riken.jp}

\author[0000-0002-4231-8717]{Gilles Ferrand}
\newcommand{\ABBL}{Astrophysical Big Bang Laboratory (ABBL), 
RIKEN Cluster for Pioneering Research, \\
Wak\={o}, Saitama, 351-0198 Japan}
\newcommand{\iTHEMS}{RIKEN Interdisciplinary Theoretical and Mathematical Sciences Program (iTHEMS), \\ 
Wak\={o}, Saitama, 351-0198 Japan}
\affiliation{\ABBL}
\affiliation{\iTHEMS}
\author[0000-0002-3222-9059]{Donald~C. Warren}
\affiliation{\iTHEMS}
\author[0000-0002-0603-918X]{Masaomi Ono}
\affiliation{\ABBL}
\affiliation{\iTHEMS}
\author[0000-0002-7025-284X]{Shigehiro Nagataki}
\affiliation{\ABBL}
\affiliation{\iTHEMS}
\author[0000-0002-4460-0097]{Friedrich~K.\ R{\"o}pke}
\newcommand{\ITA}{Zentrum f{\"u}r Astronomie der Universit{\"a}t Heidelberg, Institut f{\"u}r Theoretische Astrophysik, \\
Philosophenweg 12, 69120 Heidelberg,Germany}
\newcommand{\HITS}{Heidelberger Institut f{\"u}r Theoretische Studien, \\
Schloss-Wolfsbrunnenweg 35, 69118 Heidelberg, Germany}
\affiliation{\ITA}
\affiliation{\HITS}
\author[0000-0002-5044-2988]{Ivo~R.\ Seitenzahl}
\affiliation{School of Science, University of New South Wales, Australian Defence Force Academy, \\
Canberra, ACT 2600, Australia}
\author{Florian Lach}
\affiliation{\ITA}
\affiliation{\HITS}
\author[0000-0003-4505-8479]{Hiroyoshi Iwasaki}
\affiliation{Graduate School of Artificial Intelligence and Science, Rikkyo University, \\
3-34-1 Nishi Ikebukuro, Toshima-ku, Tokyo 171-8501, Japan}
\affiliation{Galaxies, Inc., \\
1-1-11 Minami-Ikebukuro, Toshima-ku, Tokyo, 171-0022, Japan}
\author[0000-0001-9267-1693]{Toshiki Sato}
\affiliation{High Energy Astrophysics Laboratory, 
RIKEN Cluster for Pioneering Research, \\
Wak\={o}, Saitama, 351-0198 Japan}
\affiliation{NASA, Goddard Space Flight Center, \\
8800 Greenbelt Road, Greenbelt, MD 20771, USA}
\affiliation{Department of Physics, University of Maryland Baltimore County, \\
1000 Hilltop Circle, Baltimore, MD 21250, USA\\
\vspace{8cm}}

\accepted{by ApJ}

\begin{abstract}
Progress in the three-dimensional modeling of supernovae (SN) prompts us to revisit the supernova remnant (SNR) phase. We continue our study of the imprint of a thermonuclear explosion on the SNR it produces, that we started with a delayed-detonation model of a Chandrasekhar-mass white dwarf. Here we compare two different types of explosion models, each with two variants: two delayed detonation models (N100ddt, N5ddt) and two pure deflagration models (N100def, N5def), where the N number parametrizes the ignition. The output of each SN simulation is used as input of a SNR simulation carried on until 500~yr after the explosion. While all SNR models become more spherical over time and overall display the theoretical structure expected for a young SNR, clear differences are visible amongst the models, depending on the geometry of the ignition and on the presence or not of detonation fronts. Compared to N100 models, N5 models have a strong dipole component, and produce asymmetric remnants. N5def produces a regular-looking, but offset remnant, while N5ddt produces a two-sided remnant. Pure deflagration models exhibit specific traits: a central over-density, because of the incomplete explosion, and a network of seam lines across the surface, boundaries between burning cells. Signatures from the SN dominate the morphology of the SNR up to 100~yr to 300~yr after the explosion, depending on the model, and are still measurable at 500~yr, which may provide a way of testing explosion models.
\end{abstract}


\section{Introduction}
\label{sec:intro}

Supernova remnants (SNRs) are sites of particle acceleration and inject energy and turbulence in the interstellar medium (ISM) \citep{Reynolds2017DynamicalRemnants, Vink2017X-RayRemnants}. As such, their evolution is of fundamental interest to many fields of astrophysics. Another reason to study the formation of SNRs is that they bear imprints of the stellar explosions that formed them. In the present study, we focus on remnants of Type Ia supernovae (SNe~Ia) (see \citealt{Maguire2017TypeSupernovae} for a review of their observational characteristics). SNe~Ia are essential for nucleosynthesis processes: they synthesize a significant fraction of the iron in the Universe and substantially contribute to the production of other chemical elements \citep{Seitenzahl2017NucleosynthesisSupernovae}. What placed SNe~Ia into the focus of interest recently, however, was their application as distance indicators to map out the geometry of the Universe with the resulting discovery of accelerated cosmic expansion \citep{Riess1998ObservationalConstant, Perlmutter1999MeasurementsSupernovae}.

Despite their astrophysical importance, the mechanism of SNe~Ia remains enigmatic. The basic scenario is the thermonuclear explosion of a white dwarf (WD), but important details are still unclear: What are the progenitor systems? How is the explosion triggered? How does the thermonuclear combustion front propagate through the WD material? A review of theoretical models for such thermonuclear SNe is given in \citet{Hillebrandt2013TowardsObservations}, reviews on their progenitors are \citet{Wang2012ProgenitorsSupernovae} and \citet{Ruiter2020TypeOrigin}. One of the potential scenarios is a near Chandrasekhar-mass WD accreting matter from a companion star (main sequence or evolved, non-degenerate star) until it reaches densities sufficiently high that nuclear reactions ignite in its centre \citep{Nomoto1984AccretingSupernovae}. Explosion is also possible for a sub-Chandrasekhar-mass WD, via the double detonation mechanism (accretion triggers a detonation on the surface, that eventually triggers a detonation in the core, \citealt{Nomoto1982AccretingSupernovae}), or via the violent merger of two WDs (a~double-degenerate scenario, \citealt{Pakmor2012NormalBinaries}). These two aspects may be combined \citep{Pakmor2013Helium-ignitedSupernovae}, which is known as the dynamically driven double-degenerate double-detonation scenario \citep[D6,][]{Shen2018ThreeSupernovae}. Another question is how the combustion front runs through the star: as a subsonic deflagration, a supersonic detonation, or a combination of both \citep{Ropke2017CombustionExplosions}. For the case of a Chandrasekhar-mass progenitor WD, in order to reproduce a normal SN~Ia the explosion has to start as a deflagration (accelerated by turbulent motions due to hydrodynamic instabilities) to pre-expand parts of the high-density material thus enabling the synthesis of the intermediate mass elements. The deflagration may be followed by a detonation, triggered e.g. via a spontaneous deflagration-to-detonation transition \citep[DDT,][]{Blinnikov1986DevelopmentStars, Khokhlov1991DelayedSupernovae}, the convergence of flows like in the gravitationally-confined detonation \citep[GCD,][]{Plewa2004TypeDetonation}, or some other process (see \citealt{Poludnenko2019ASupernovae} for a recently proposed mechanism). Pure detonations of a Chandrasekhar-mass WD are not good models for normal SNe~Ia because they produce too much Fe-group elements, mostly $^{56}$Ni, as almost the whole star goes through Si-burning.
The outcome of explosions of near Chandrasekhar-mass WDs remains uncertain, and motivates on-going numerical works.
The different cases for the combustion produce ejecta with different morphology and composition (e.g. \citet{Seitenzahl2013Three-dimensionalSupernovae} and \citet{Fink2014Three-dimensionalSupernovae}). 

Our work belongs to the line of studies seeking to use the (young) supernova remnant as a probe of the explosion mechanism \citep{Milisavljevic2017TheConnection,Patnaude2017SupernovaProgenitors}. The~SNR is formed by the interaction between the supersonic ejecta and the ambient medium. Young SNRs are characterized by a shell-like structure, bounded by two shocks: reverse shock (RS) in the ejecta and forward shock (FS) in the ambient medium, that process kinetic energy into thermal energy (heating of the plasma) and non-thermal energy (magnetic field amplification and particle acceleration). The interface between the ejecta and the ambient medium, called the contact discontinuity (CD), is subject to the Rayleigh-Taylor instability (RTI), which shapes the morphology of young SNRs observed in radio and in X-rays \citep{Chevalier1992HydrodynamicWaves}.
Remnants from thermonuclear SNe can be distinguished from remnants from core-collapse SNe with different methods. Ideally one would want to observe the supernova itself, this is still possible for ancient events thanks to light echoes (see for example \citet{Krause2008TychoSpectrum} and \citet{Rest2008Scattered-Light1572} for Tycho's SNR). The presence of a remaining compact object may reveal the explosion class: a neutron star is evidence for a core-collapse SN, while an unusual white dwarf may be the product of a failed thermonuclear SN. Another clue is the kind of environment: core-collapse SNe are associated with massive stars, and their forming regions. X-ray diagnostics play a key role, spectral diagnostics include metal abundances \citep[e.g.][]{Reynolds2007AInteraction} and the position of the Fe-K line centroid \citep{Yamaguchi2014DiscriminatingEmission}, morphological diagnostics are available as well \citep{Lopez2009TypingMorphologies}. Can we do more, and distinguish different kinds of thermonuclear SNe? Lines of investigation include: the search for surviving companions \citep{Ruiz2019SurvivingObservations}, and their circumstellar medium; detailed composition studies that allow us to conclude on characteristic nucleosynthesis effects, e.g. looking for neutronized species like $^{58}$Ni and $^{55}$Mn \citep{Seitenzahl2013SolarProgenitors,Yamaguchi2015AManganese}; morphological studies \citep[e.g.][]{Yamaguchi2017TheRemnant,Williams2017TheRemnant,Sato2020ARemnant}; and the detection of coronal lines of the non-radiatively shocked ejecta \citep{Seitenzahl2019OpticalSupernovae}.

This work focuses on the SNR morphology, with a view on the existing and upcoming X-ray observations. In \citet[][hereafter Paper~I]{Ferrand2019FromExplosion} we presented the 3D evolution of the early SNR phase from one 3D model of a SN explosion, a~delayed detonation of a Chandrasekhar-mass WD. We investigated until what age the SN explosion dynamics leaves an imprint on the structure of the SNR. In this paper, still considering the explosion of a Chandrasekhar-mass WD, we compare four different SN models along two aspects: the ignition asymmetry, as parametrized by the number of ignition kernels, and the burning front propagation, either a deflagration-to-detonation \citep{Seitenzahl2013Three-dimensionalSupernovae} or a pure deflagration \citep{Fink2014Three-dimensionalSupernovae}. In~Section~\ref{sec:method} we present the SN models used, and the subsequent SNR simulation. In~Section~\ref{sec:results} we present our results on the SNR morphology, illustrated by two kinds of plots: plots in slab geometry that show the SNR as observed from a given line of sight, and plots in spherical geometry that show the entire surface of the SNR. We discuss these results with respect to observations in Section~\ref{sec:discussion}. Finally in Section~\ref{sec:conclusion} we summarize our study and present our perspectives.
\\~

\section{Method}
\label{sec:method}

We are relying on published SN models, thereby extending the work by \citet{Seitenzahl2013Three-dimensionalSupernovae} and \citet{Fink2014Three-dimensionalSupernovae}. The main properties of the models are summarized in Section~\ref{sec:method-SN}. We have simulated the early SNR phase, as was done in Paper~I. The numerical method and the analysis strategy are summarized in Section~\ref{sec:method-SNR}.

\subsection{SN models}
\label{sec:method-SN}

\afterpage{

\begin{deluxetable}{lllll}
\tablehead{\colhead{} & \colhead{$E_{\rm kin}$ (erg)} & \colhead{$M_{\rm ej}$ ($M_\odot$)} & \colhead{$v_{\rm max}$ (km.s$^{-1}$)} & \colhead{composition}}
\startdata
N100ddt & $1.43\times10^{51}$ & $1.40$ & $28,700$ & 60\% IGE, 30\% IME \\
N5ddt   & $1.55\times10^{51}$ & $1.40$ & $27,900$ & 80\% IGE, 10\% IME \\
N100def & $6.15\times10^{50}$ & $1.35$ & $14,100$ & 40\% IGE, 10\% IME \\
N5def   & $1.35\times10^{50}$ & $0.37$ & $13,800$ & 54\% IGE, ~5\% IME \\
\enddata
\caption{\label{tab:SN_models}
Summary data for the explosion models: explosion's (asymptotic) kinetic energy $E_{\rm kin}$, ejected mass $M_{\rm ej}$, maximum ejecta velocity $v_{\rm max}$ (100~s after the ignition), and global composition of the ejecta (as a fraction of the unbound mass) where IGE = iron group elements and IME = intermediate group elements (see detailed yields in \citet{Seitenzahl2013Three-dimensionalSupernovae} for the ddt models and in \citet{Fink2014Three-dimensionalSupernovae} for the def models).}
\end{deluxetable}

\begin{figure}
\centering
\includegraphics[width=1\textwidth]{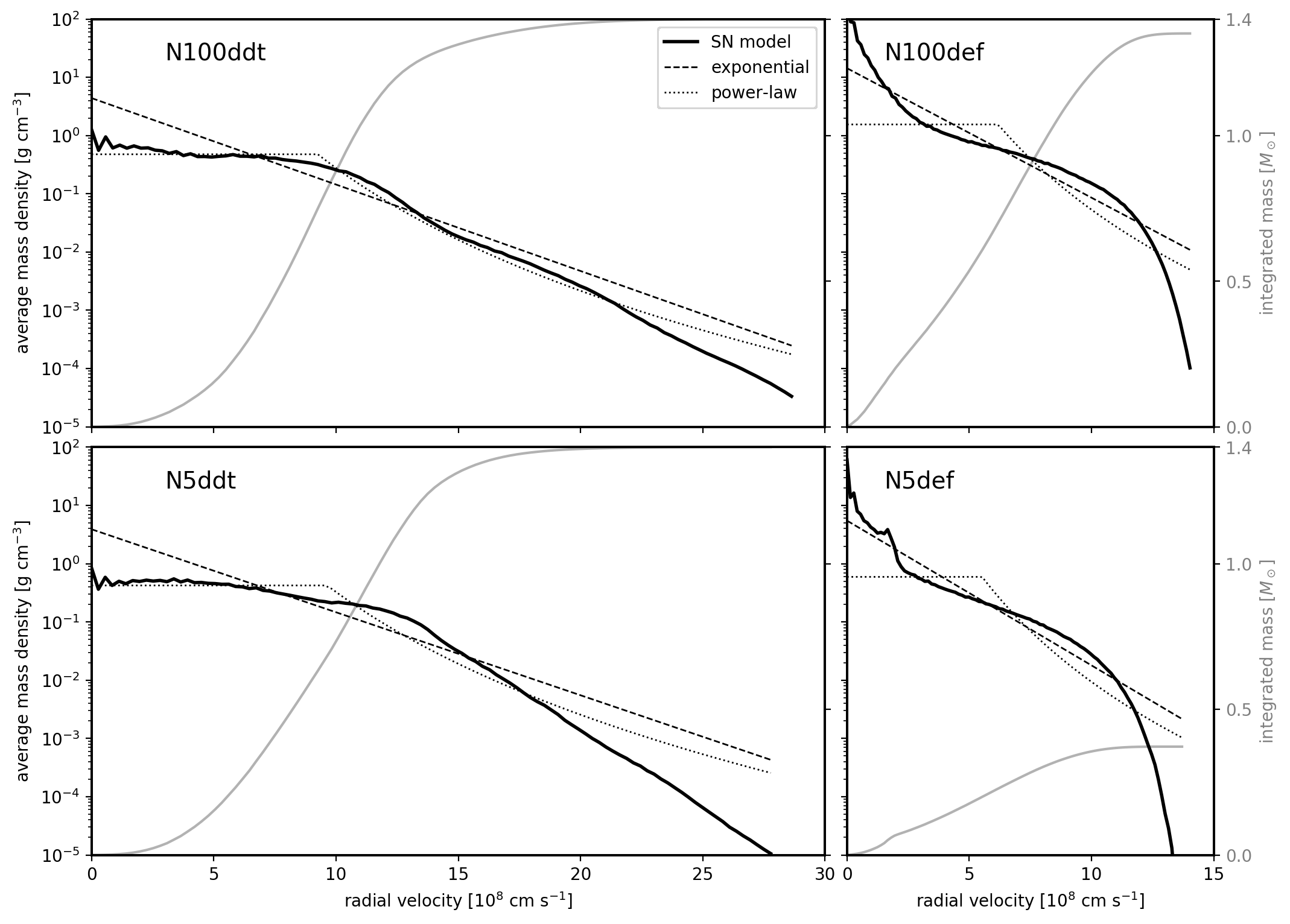}
\caption{\label{fig:rho-1D}
Average radial profile of the mass density of the SN models at t = 100~s (thick solid line), as a function of radial velocity. Note that the density range is the same for all plots, while the velocity coordinate is twice as extended for ddt models vs.\ def models. Thinner lines are analytical profiles with the same mass and energy: an exponential law (dashed) and a truncated power law of index $n=7$ (dotted). The gray curve shows the integrated mass, read on the rightmost axis.}
\vspace{15mm}
\end{figure}
}

Our model's basic properties are given in Table~\ref{tab:SN_models}. All models consider the explosion of a single accreting WD (single degenerate model), of Chandrasekhar mass $M = 1.4 M_\odot$, initial radius $R = 1.96\times10^8$~cm, and homogeneous composition of $^{12}$C and $^{16}$O in equal parts by mass. The zero-age main-sequence progenitor is assumed to be of solar metallicity, with electron fraction $Y_e=0.49886$.\footnote{In the explosion simulation, $Y_e$ is followed independently of the simplified chemical composition.} 
The four SN models we have selected cover two explosion mechanisms: there are two delayed detonation models (ddt) from \citet{Seitenzahl2013Three-dimensionalSupernovae} and two deflagration models (def) from \citet{Fink2014Three-dimensionalSupernovae}.\footnote{Note that in \citet{Seitenzahl2013Three-dimensionalSupernovae} N100ddt and N5ddt models are just called N100 and N5, in \citet{Fink2014Three-dimensionalSupernovae} the suffix ``def" was added to the pure deflagration versions, here we are adding the suffix ``ddt" to the original deflagration-to-detonation versions, for balance and clarity.} We also compare two numerical setups for the central ignition: two models have many ignition points (N100), leading to rather spherical ejecta, while two models have a few ignition points (N5), leading to more asymmetric ejecta. We note that the N~numbers are not to be taken too literally, rather, the ignition spots should be thought of as a convenient parametrization of the initial asymmetry and filling factor of the flame surface. The ignition physics in such highly turbulent flows is not well understood; the most advanced simulations to date find that multiple ignition is unlikely and off-center ignition is likely \citep{Zingale2009LowConvection, Nonaka2012High-ResolutionRefinement}. Our ambition here is not to create a complete mapping between such ignition models and their parametrizations. In this paper, we focus on assessing the impact of existing parametrized ignition models on the eventual morphology of the supernova remnants.

In the grid of ddt models from \citet{Seitenzahl2013Three-dimensionalSupernovae}, N100ddt produces a SN of normal brightness from the synthesized mass of $^{56}{\rm Ni} \simeq 0.6 M_\odot$, and it reproduces well key spectral characteristics of normal SNe~Ia \citep{Sim2013SyntheticSupernovae}. This is thus the model we first picked, in Paper~I, to make simulations of the SNR phase. It~is a rather symmetric explosion \citep{Bulla2016PredictingSupernovae}. N5ddt was selected next, because having less ignition kernels it is more asymmetric. The fewer ignition kernels, the weaker the initial deflagration, and the more fuel is processed by the detonation, therefore N5ddt is also brighter than N100ddt. Both N100ddt and N5ddt were considered in \citet{Williams2017TheRemnant} for the interpretation of the ejecta distribution in Tycho's SNR. The N3ddt model, which is similar to N5ddt, was discussed in \citet{Borkowski2013SupernovaG1.9+0.3} for the SNR G1.9+0.3. From \citet{Sim2013SyntheticSupernovae} the best matches for the low N ddt models are with 1991T-like SNe~Ia, although agreement is not great in details. 

The models in the grid of \citet{Fink2014Three-dimensionalSupernovae} are the same as in \citet{Seitenzahl2013Three-dimensionalSupernovae}, the only difference being that the deflagration-to-detonation transition is not allowed. Whereas all our ddt models fully unbind the WD, def models are often partial explosions that leave compact remnants behind, that is, a WD with an unusual composition -- which may have been observed, see \citet{Vennes2017AnSupernova}. Such a weak explosion like N5def is thought to be a good model for SN2002cx-like SNe~Ia (\citealt{Kromer20133DSupernovae}; see also \citealt{JordanIV2012Failed-detonationCores} for a similar model) and maybe for the larger class of Type Iax SNe \citep{Bulla2020WhiteInteraction}. N5def is an asymmetric explosion that leaves behind a $\approx 1~M_\odot$ WD remnant. Compared with N5ddt, the ejected mass is 3.8 times lower while the kinetic energy is 11.5 lower, the maximum velocity reached is about halved. For completion, we also include the N100def model, although it appears that the strongest def events cannot explain any observed class of SN~Ia. Contrary to N5def, N100def is almost all unbound, just a bit less energetic than the ddt version. Compared with N100ddt, the ejected mass is almost the same, while the energy is 2.3 times lower, and the maximum velocity is again about halved.

The SN explosion simulations were made in 3D using the hydrodynamic code \textsc{LEAFS} \citep{Reinecke2001RefinedSimulations}. Each simulation was done in two steps. First a hydro run with a restricted nuclear reaction network, with only the species needed to get the energy budget right: carbon, oxygen, a proxy for intermediate-mass isotopes, nickel, and alpha particles. Then the computation of the nucleosynthesis in post-processing, using one million tracer particles and a full nuclear reaction network with 384 isotopes of elements up to $Z=32$.
Each model was simulated until time 100~s, well after the end of nuclear reactions, and when the ejecta are observed to have settled into the self-similar free expansion phase. 
The final result has been remapped to Cartesian grids of resolution $200^3$.

Figure~\ref{fig:rho-1D} shows the angularly-averaged radial profile of the density, for the four models. The averaging is done from the explosion centre, from which everything expands. The average radial density profiles of the ddt and def models are different. The average density of ddt models is rather uniform in the core and then decreases exponentially, whereas for def models it peaks towards the centre (at low velocities) and falls down more rapidly at the edge. 
The obtained profiles are compared with simple analytical profiles commonly used in the SNR community: a power-law profile \citep[e.g.][]{Decourchelle1994ModelingRemnant}, and an exponential profile \citep{Dwarkadas1998InteractionSurroundings}, none of which are a very good fit. 
Anyway, such a 1D density profile does not capture the complicated structure of the ejecta, that stems from the dynamics of the deflagration and/or detonation fronts. As shown in Paper~I, when reducing the full 3D profile to a single radial profile one does not get the same SNR evolution.

\subsection{SNR simulations}
\label{sec:method-SNR}

The output of the SN simulations described in the previous section were used as input for the SNR simulations described in this section. See Paper~I for details. For the def models, the bound remnant has been excised for the simulation of the SNR phase. As explained in \citet{Fink2014Three-dimensionalSupernovae}, to determine the unbound parts we calculated the asymptotic specific kinetic energy $\epsilon_{\rm kin,a} = \epsilon_{\rm kin} + \epsilon_{\rm grav}$, the condition to be unbound is $\epsilon_{\rm kin,a}>0$, then matter will approach velocity $v_{\rm a} = \sqrt{2\epsilon_{\rm kin,a}}$. During the initial self-similar free expansion evolution phase, $r \propto t$ and $\rho \propto t^{-3}$. To save on computing time, the ejecta profiles are re-scaled from $t=100$~s to $t=1$~day, and the hydro simulation starts at 1~day. It~is observed that the assumption of self-similarity is actually reasonable up to about a year of evolution. We note that we are not concerned here with the SN phase per se, and the modeling of the light curve, although we checked the effect of early energy deposition from radioactive decay on the dynamics (see Section~\ref{sec:discussion-other}). 
The ambient density is assumed to be uniform for simplicity, and of low density $n_\mathrm{ISM}=0.1~{\rm cm}^{-1}$ to have overall dynamics similar to Tycho's SNR. Besides this value, the evolution of the SNR depends on the explosion energy $E_{\rm SN}$ and ejecta mass $M_{\rm ej}$ \citep{Truelove1999EvolutionRemnants}. From these three quantities one can define three characteristic scales, we are using here the same definitions as in \citet{Dwarkadas1998InteractionSurroundings} and \citet{Warren2013Three-dimensionalSNRs}:
\begin{eqnarray}
r_\mathrm{ch} &=& \left(\frac{3 M_\mathrm{ej}}{4 \pi \rho_\mathrm{ISM}}\right)^{1/3} , \label{eq:r_ch} \\
u_\mathrm{ch} &=& \left(\frac{2E_\mathrm{SN}}{M_\mathrm{ej}}\right)^{1/2} , \label{eq:u_ch} \\
t_\mathrm{ch} &=& \frac{r_\mathrm{ch}}{u_\mathrm{ch}} . \label{eq:t_ch}
\end{eqnarray}
The hydrodynamic evolution of the SNR is computed up to 500 yr after the explosion (about the age at which we observe Tycho's SNR).

The SNR hydro simulation is performed using a custom version of the hydrodynamics code \textsc{RAMSES} \citep{Teyssier2002CosmologicalRAMSES}. 
Since we are interested in the morphology of the remnant, we conduct simulations in 3D. They are performed on a Cartesian grid, of resolution $256^3$. The SN grid is loaded at 1:1, leaving room for the blast wave developing ahead of the ejecta. The grid is co-expanding with the remnant, so that the physical size of the box is changing over time, while the relative resolution of the cells remains constant. 
Our purpose is to study the morphology of the remnant, independent of its physical size (which for a given SNR at a given age requires knowledge of its distance, which is usually very uncertain). 

We focus on the shocked region, which is the part revealed by X-ray observations. To make visible its 3D structure, and to quantify its evolution, we proceed as described in Paper~I. At any given time, we extract in the simulation box the surface of the wave fronts: the reverse shock (RS) in the ejecta, the contact discontinuity (CD) between the shocked ejecta and the shocked ISM, the forward shock (FS) in the ISM. We record the radius of each surface, measured from the explosion centre, as a function of the direction $(\theta,\phi)$. These data are stored and analysed using the HEALPix package \citep{Gorski2005HEALPixSphere}. We interpret these radii (or functions thereof) as functions $R(\theta,\phi)$ on the sphere, that we expand in spherical harmonics $Y_{\ell}^{m}(\theta,\phi)$, to obtain their angular power spectrum $C_\ell$. The complete set of plots from this analysis, for all models and all wavefronts, is available in an online repository.\footnote{DOI: \href{http://doi.org/10.5281/zenodo.4040625}{10.5281/zenodo.4040625} }
We checked that the reconstructed maps are indistinguishable from the original ones by eye, and that the residuals have negligible levels for our analysis (these ancillary plots are present in the repository). The residuals show random variations at the pixel scale, plus some large scale patterns and/or some hot spots, that may or may not have an obvious counterpart. Across the 4~models and the 500 years of evolution, the relative residuals have zero mean, standard deviation below 1\% for the CD and below 0.1\% for the shocks, and local peaks at up to 8\% for the CD and 3\% for the shocks. The angular spectra presented in Section~\ref{sec:results-healpix} capture quantitatively the most part of what can be seen from the simulation results.\footnote{In order to better capture localized structures, a more flexible approach may be wavelet analysis. For applications in spherical geometry, see \citet{Starck2010SparseDiversity} and \citet{Marinucci2011RandomApplications}.} They allow us to see the different evolution of SN-phase modes and SNR-phase modes (see also Paper~I) and to make comparisons between the different models (new to this paper).

Furthermore, each SNR simulation is made with two setups: one using the actual SN profile (labelled 3Di), and one using an angularly-averaged version of it, so effectively one-dimensional, depending on radius only (labelled 1Di). The results from the 1Di simulations will not be all shown in this paper, but are used as a reference to interpret the results of the 3Di simulations, as well as to assess the possible presence of numerical artefacts (see Paper~I for a systematic comparison for the N100ddt model).
\\~

\section{Results}
\label{sec:results}

\begin{figure}[p]
\centering
\includegraphics[width=1\textwidth]{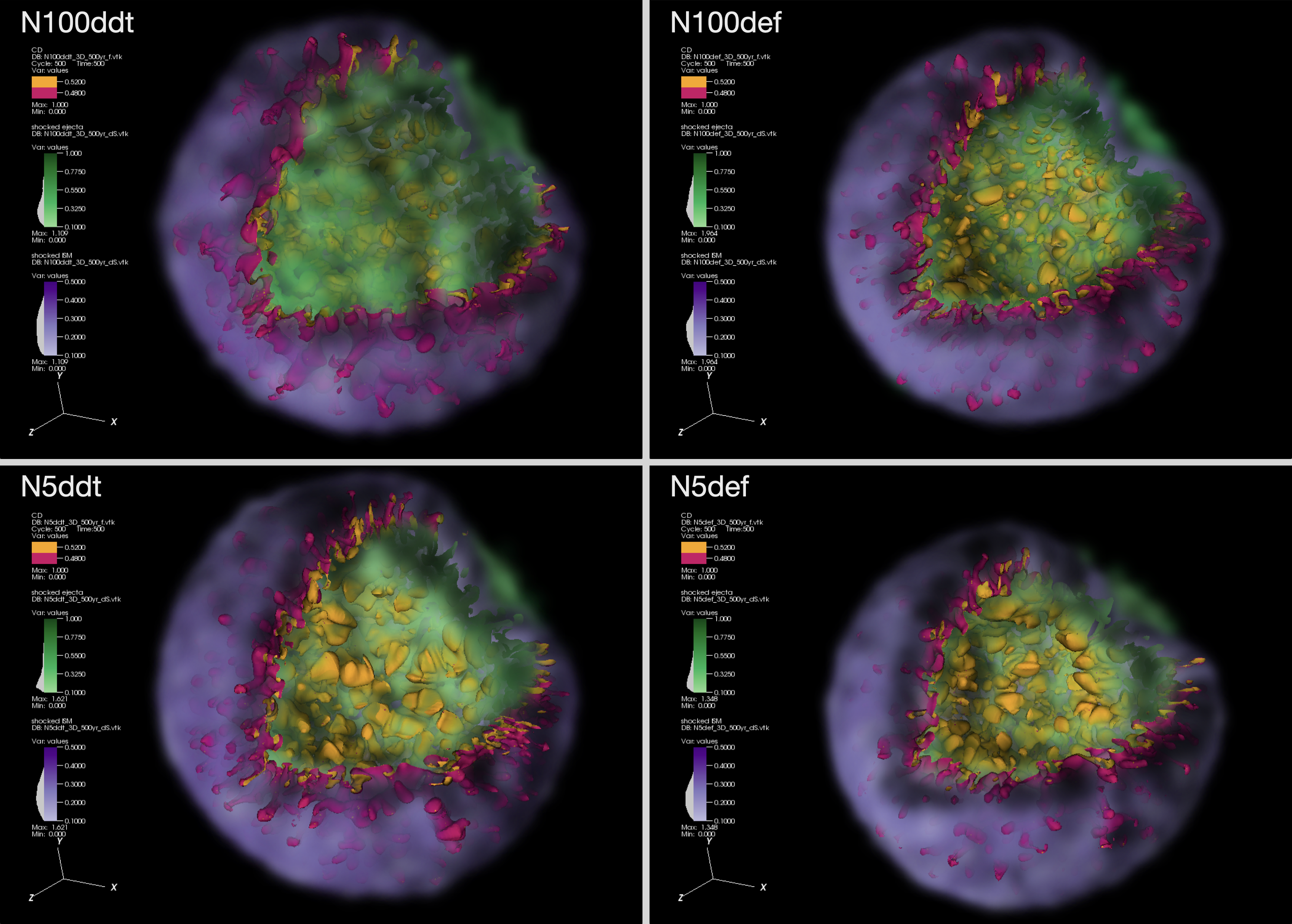}
\caption{3D view of the SNR models at 500 years. The shocked ISM (located between the FS and the CD) is volume-rendered in purple hues. The shocked ejecta (located between the RS and the CD) is volume-rendered in green hues. The interface between the two (the CD) is highlighted with a solid iso-contour, its inner side is coloured yellow while its outer side is coloured red.  The layer of ISM, and the CD contour, have been removed over a quarter of the remnant to reveal its interior. The physical radius of the SNR (measured from the explosion centre) at $t=500$~yr is N100ddt: $5.6\pm0.1$~pc, N5ddt: $5.8\pm0.2$~pc, N100def: $4.6\pm0.1$~pc, N5def: $3.6\pm0.3$~pc.} 
\label{fig:3D-view}
\end{figure}

In Figure~\ref{fig:3D-view} we show 3D views of the four SNR models, 500~yr after the explosion. By that age all models assume a roughly spherical shape. The young SNR exhibits the typical two-shock structure of the ejecta-dominated phase. The Sedov-Taylor timescale $t_{\rm ST}$ (the age at which as much ISM mass is swept-up as the ejected mass) happens to be around 500~yr for N100dt, N5dddt and N5def models, a bit larger, around 700~yr, for N100def; after a few $t_{\rm ST}$ all our SNRs will converge to the Sedov-Taylor solution.
Even though differences between the models are visible at 500~yr, they are not striking when looking at the global 3D morphology (and not easy to investigate without interactivity). In the following, we present a selection of 2D plots (in slab geometry and in spherical geometry) that reveal the structure of the SNR, and guide the reader through the distinctive features of the different models (N100 vs.\ N5 and ddt vs.\ def). 
In Section~\ref{sec:results-maps} we use density maps (slices and projections) which are easy to interpret but can only show one section along one direction at a time, in Section~\ref{sec:results-healpix} we use spherical projections which allow us to analyse the entire surface of the SNR.

\subsection{2D density maps}
\label{sec:results-maps}

\paragraph{Slices.}

In Figures~\ref{fig:map-cut-N100ddt},\ref{fig:map-cut-N5ddt},\ref{fig:map-cut-N100def},\ref{fig:map-cut-N5def} we show for each model slices of the mass density, to show the inner structure of the remnant. On each plot we show a slice in the mid-plane along three directions (the principal axes of the grid $x$, $y$, $z$), to appreciate the morphological variations, and a slice at three different ages, to appreciate the time evolution: at 1~yr the morphology is still very similar to the initial conditions, at 100~yr the characteristic shell of the SNR phase is visible while the imprint of the SN is still clear, at 500~yr the shell-like structure is on the way of regularization. Movies from 1~yr to 500~yr in steps of 1~yr are also available. 
In all cases, we can see the competition between two effects shaping the young SNR: the progressive regularization of the SN ejecta, and the growth of the RTI. After a few hundreds of years, all the SN models, despite differences in energy explosion and ejecta distribution, tend to the standard shell structure expected for a young SNR.

\def\widthmap{1\textwidth}

\begin{figure}[p]
\includegraphics[width=\widthmap]{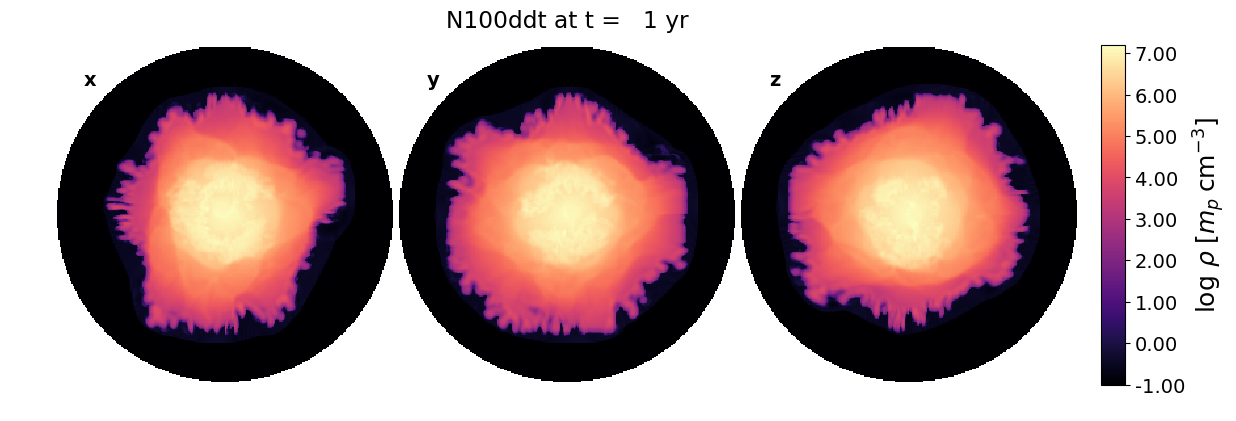}
\includegraphics[width=\widthmap]{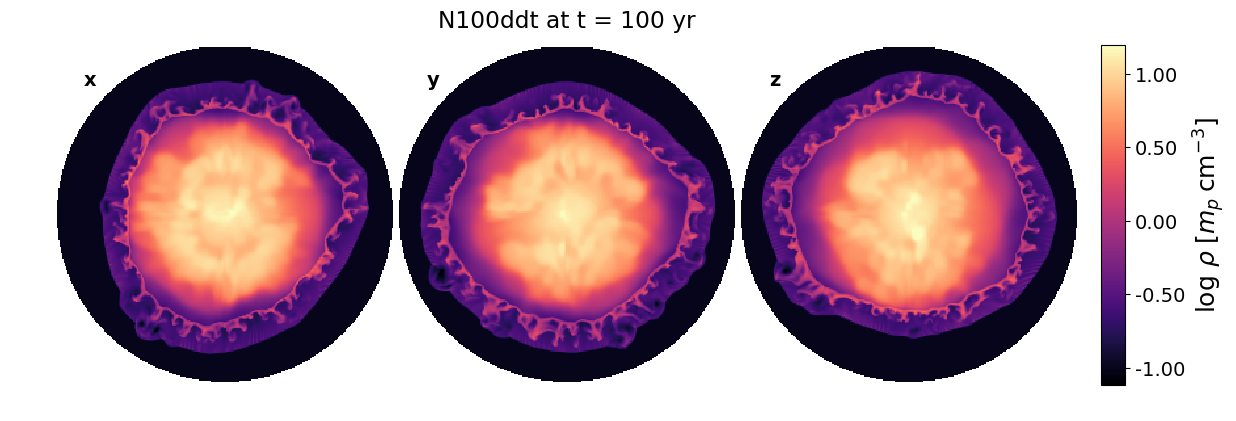}
\includegraphics[width=\widthmap]{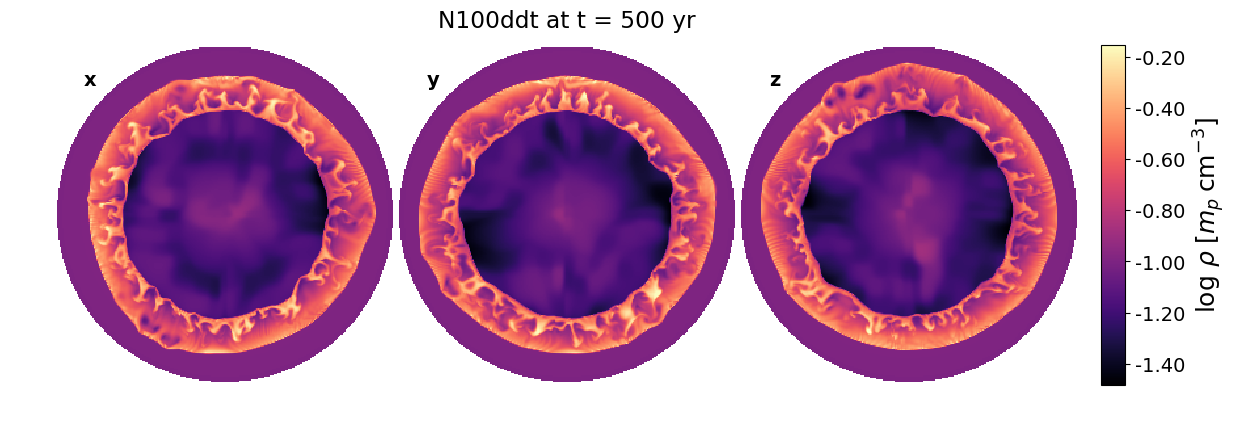}
\caption{Slices of density for N100ddt, in the mid-plane along three different axes (principal axes x, y, z, of the simulation box), at three different times: 1~yr, 100~yr, 500~yr (note that the colour scale is adjusted at each time to match the density range). The size of the box is $0.0857$~pc at 1~yr, $4.80$~pc at 100~yr, $13.4$~pc at 500~yr. An animation over time from 1 to 500~yr in steps of 1~yr is available (movie duration: 10~s). 
\label{fig:map-cut-N100ddt}}
\end{figure}

\begin{figure}[p]
\includegraphics[width=\widthmap]{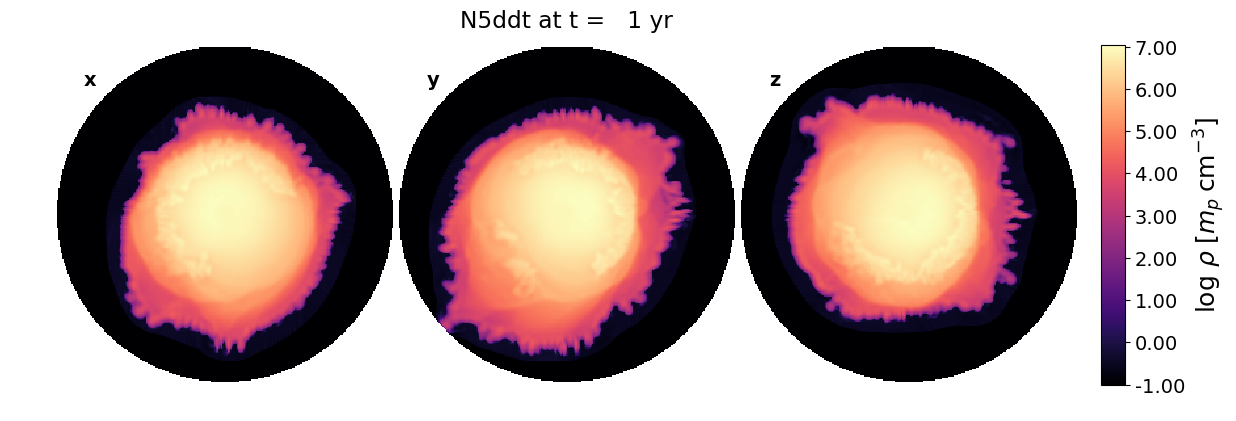}
\includegraphics[width=\widthmap]{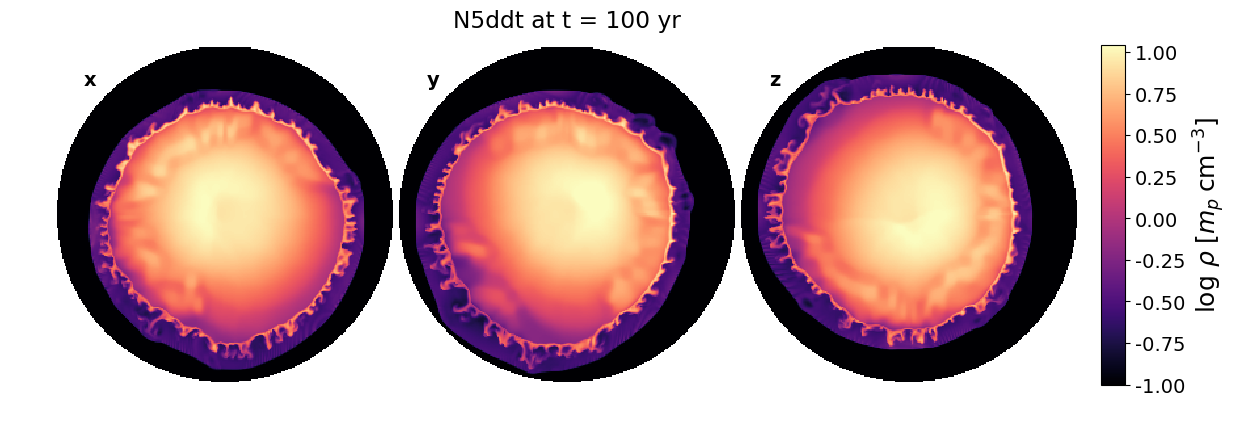}
\includegraphics[width=\widthmap]{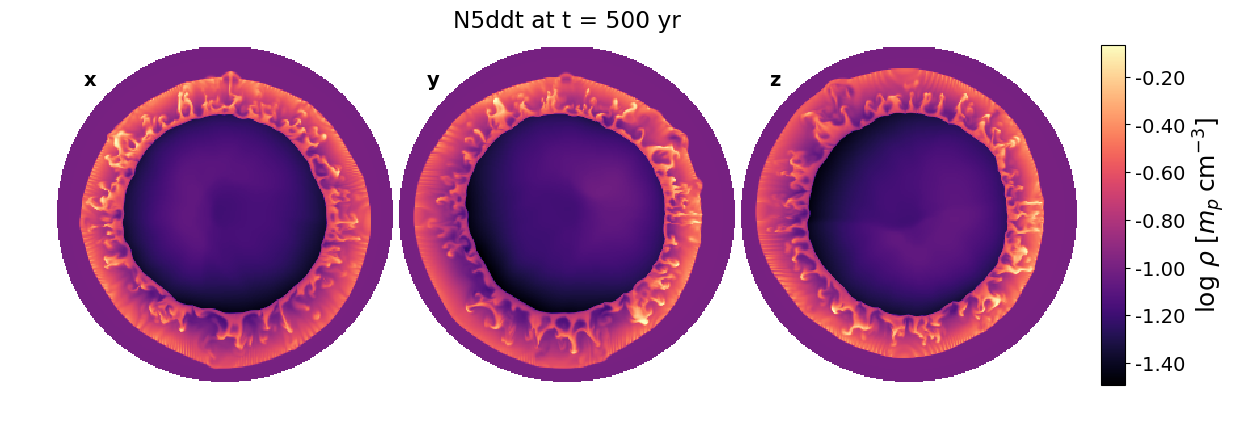}
\caption{Slices of density for N5ddt. The size of the box is $0.0704$~pc at 1~yr, $4.32$~pc at 100~yr, $13.6$~pc at 500~yr. An animation over time from 1 to 500~yr in steps of 1~yr is available (movie duration: 10~s).
\label{fig:map-cut-N5ddt}}
\end{figure}

\begin{figure}[p]
\includegraphics[width=\widthmap]{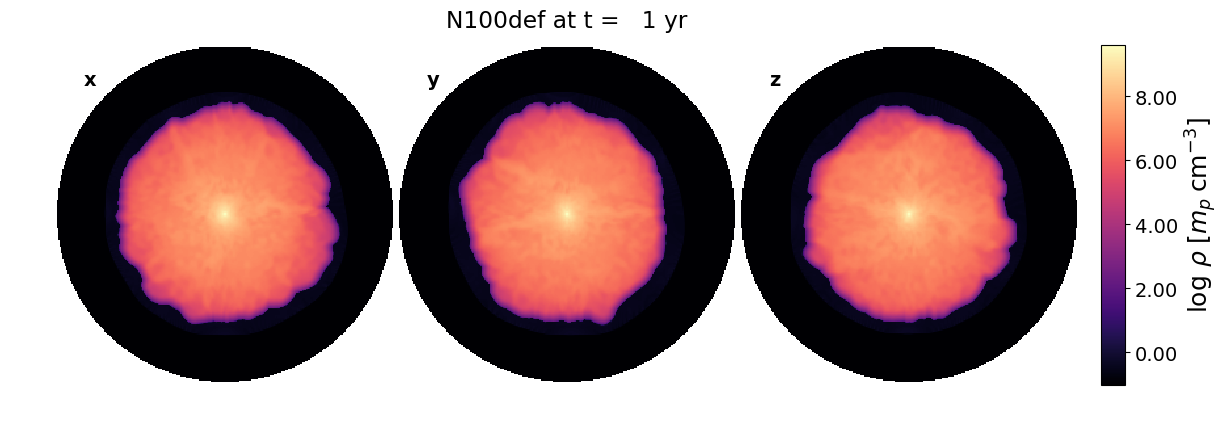}
\includegraphics[width=\widthmap]{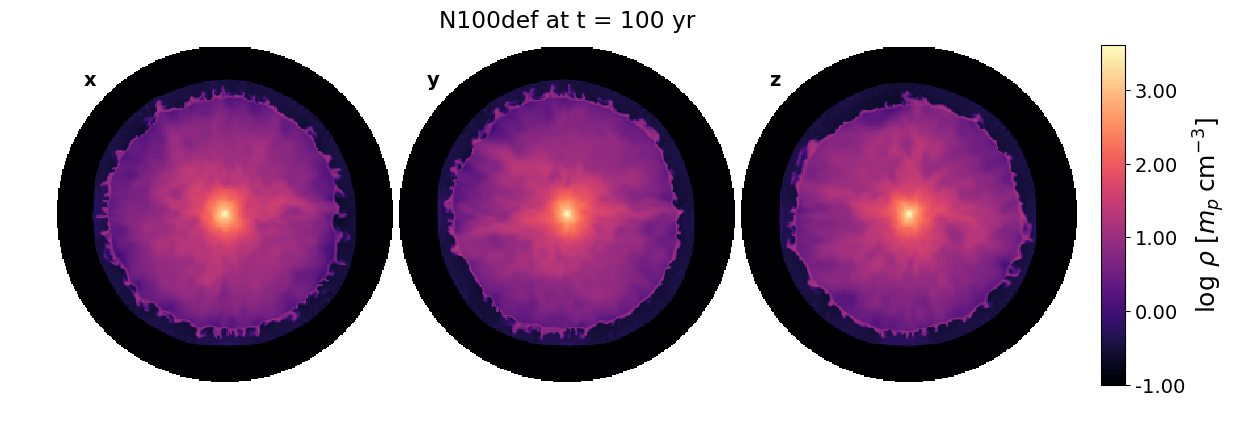}
\includegraphics[width=\widthmap]{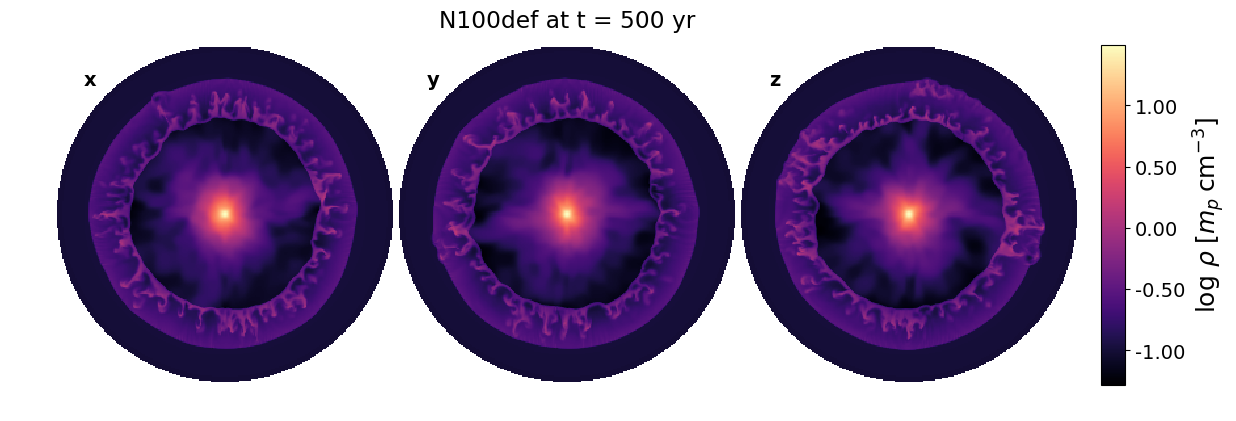}
\caption{Slices of density for N100def. The size of the box is $0.0438$~pc at 1~yr, $3.20$~pc at 100~yr, $11.6$~pc at 500~yr. An animation over time from 1 to 500~yr in steps of 1~yr is available (movie duration: 10~s).
\label{fig:map-cut-N100def}}
\end{figure}

\begin{figure}[p]
\includegraphics[width=\widthmap]{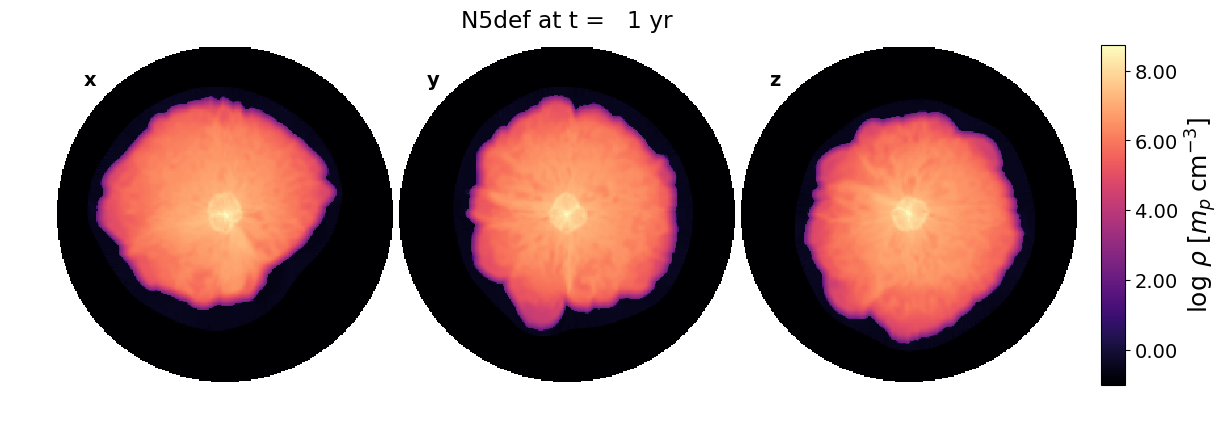}
\includegraphics[width=\widthmap]{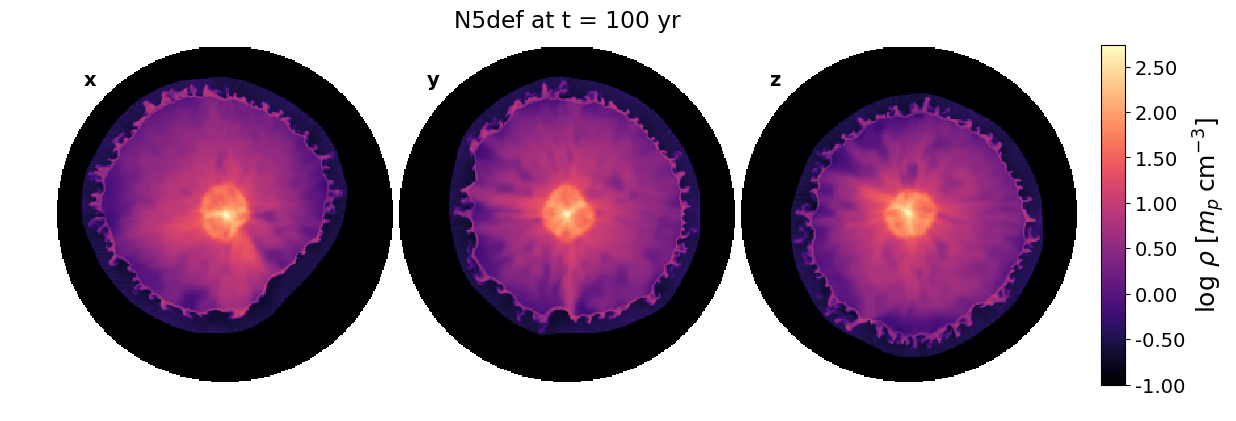}
\includegraphics[width=\widthmap]{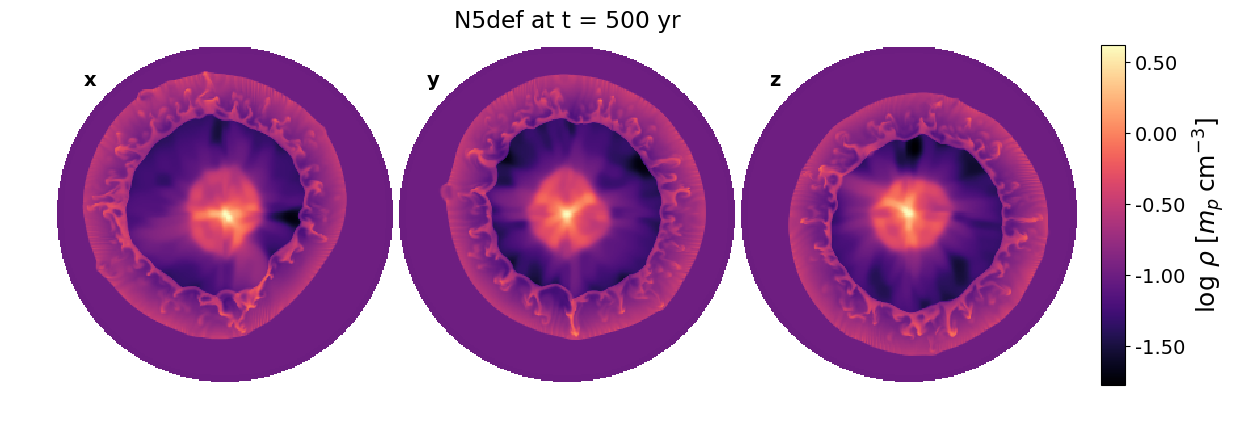}
\caption{Slices of density for N5def. The size of the box is $0.0384$~pc at 1~yr, $2.82$~pc at 100~yr, $9.36$~pc at 500~yr. An animation over time from 1 to 500~yr in steps of 1~yr is available (movie duration: 10~s).
\label{fig:map-cut-N5def}}
\end{figure}

For N100ddt, as previously reported, by 500~yr the SNR has regularized in a rather spherical shell, although with some disturbances compared with the case of smooth initial conditions (see the detailed comparison of the 1Di and 3Di setups in Paper~I). The ejecta initially have a shape that, although complicated, does not exhibit a~preferred direction (this is quantified in Section~\ref{sec:results-healpix}), as a result the shell structure at 500~yr looks statistically similar when observed at different angles. 
The situation is different for N5ddt. This model initially has an elongated structure, which over time leads to a two-sided SNR, with one half of the shell being almost twice as thick as the other half. Looking more carefully at the time evolution, we see that the FS is located at different radii from the very beginning, with the layer of swept-up ISM initially growing uniformly. The more elongated parts of the ejecta are also the less dense, and so the parts where the RS travels faster inward, this is the main reason for the growing spread between the RS and the FS. The net result at 500~yr is that both the RS and the FS are nearly circular, but the RS is actually centred on the explosion centre whereas the FS is off-centred. The RT fingers grow in a similar way over the CD, and so are more likely to interact with the FS on the thin side than on the thick side. 

This asymmetry difference between the N100 and N5 models is also visible for the def cases, although it expresses itself in different ways. Because of the way the deflagration fronts propagate, def models tend to have more regular initial contours, and with a more uniform density distribution over angles. This can be quantified, in a given slice, by comparing the perimeter of the CD with the perimeter of a circle having for radius the average radius of the CD. At 1~yr this ratio is N100ddt: $5.7\pm0.5$, N5ddt: $4.8\pm0.5$, N100def: $2.2\pm0.0$, N5def: $2.6\pm0.1$. So from the start def models have a contour twice as regular as ddt models. 
As a result def models produce rather symmetric SNRs. But whereas N100def has a fairly isotropic distribution, from the start and so at all times, N5def is clearly biased on one side, leading to an off-centred SNR. Compared to N5ddt, the ejecta density being more uniform the RS travels in a more symmetric way, leading to a rather balanced shell. But now both shocks, and so the entire SNR shell, is completely off-centred w.r.t the explosion centre, at all times. 
Another remarkable property of def models is the presence of an over-density at the centre. This stems from the fact that pure deflagration models fail to fully unbind the WD. There is, at the centre of the peak, a left-over compact remnant. We remind the reader that this remnant is not included in the SNR simulations, only the unbound ejecta are taken into account; we add a word of caution that the cut may not be perfectly modelled, although the presence of low-velocity ejecta is a real feature. For the N5def model, for which only a quarter of the WD mass is ejected, we even see a shock front near the centre, which originates in the sloshing motion of the matter that is marginally unbound \citep{Fink2014Three-dimensionalSupernovae}. By 500~yr the RS is still far from these structures. We checked that at 1000~yr, the RS has just started its motion inwards (in the ISM frame), but has barely reached the central over-dense region for the def models. Semi-analytical models (assuming spherically symmetric ejecta with a power-law radial density profile), as implemented in the calculator by \citet{Leahy2017AEvolution}, predict it will take up to 3000~yr for the RS to reach the centre with our range of $E_{\rm kin}$ and $M_{\rm ej}$.
For the def models, especially N100def, we also observe dense filaments radiating from the center. Looking back at the time evolution of the N100def SN simulation, we see how these structures are formed as the deflagration develops (see Figures~\ref{fig:N100def-SN-rho} and~\ref{fig:N100def-SN-xCO} in the Appendix). These over-dense channels are created between large-scale plumes of the deflagration ashes. Being made of unburnt material, they are rich in C and O.

\paragraph{Projections.}

\begin{figure}[p]
\includegraphics[width=\widthmap]{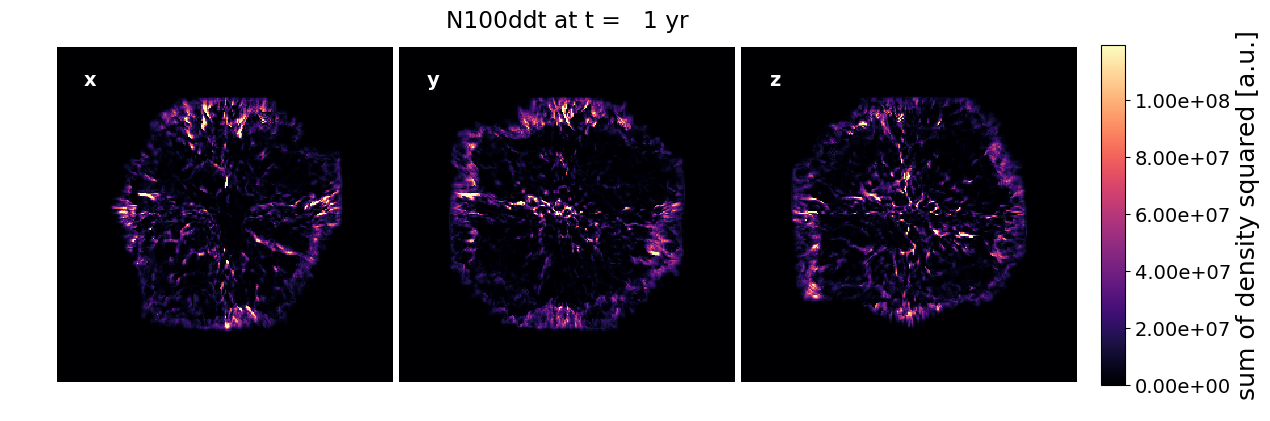}
\includegraphics[width=\widthmap]{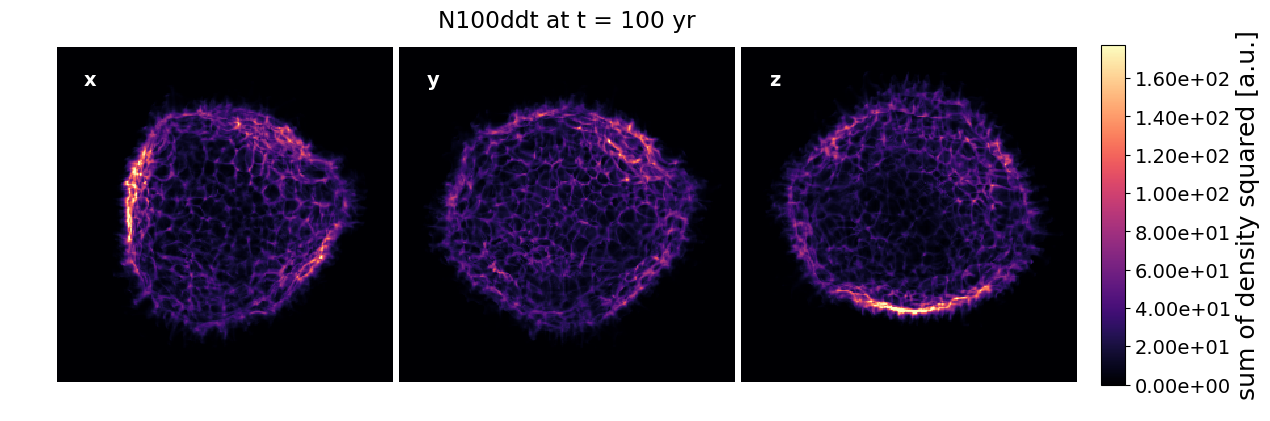}
\includegraphics[width=\widthmap]{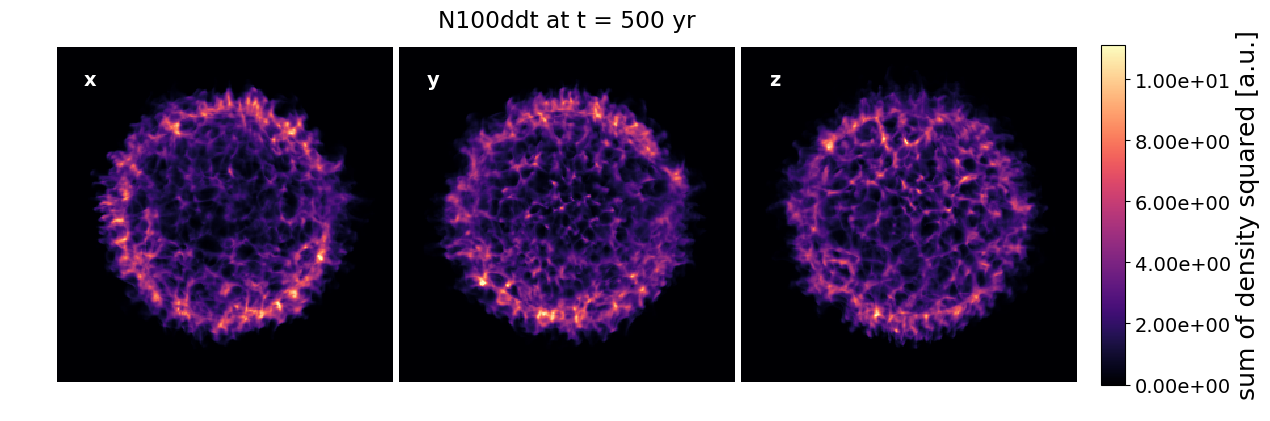}
\caption{Projected maps of density squared in the shocked region (proxy for the broad-band thermal emission) for N100ddt, along three different axes (principal axes x, y, z, of the simulation box), at three different times: 1~yr, 100~yr, 500~yr (note that the colour scale is adjusted independently at each time). The size of the box is $0.0857$~pc at 1~yr, $4.80$~pc at 100~yr, $13.4$~pc at 500~yr. An animation over time from 1 to 500~yr in steps of 1~yr is available (movie duration: 10~s).
\label{fig:map-prj-N100ddt}}
\end{figure}

\begin{figure}[p]
\includegraphics[width=\widthmap]{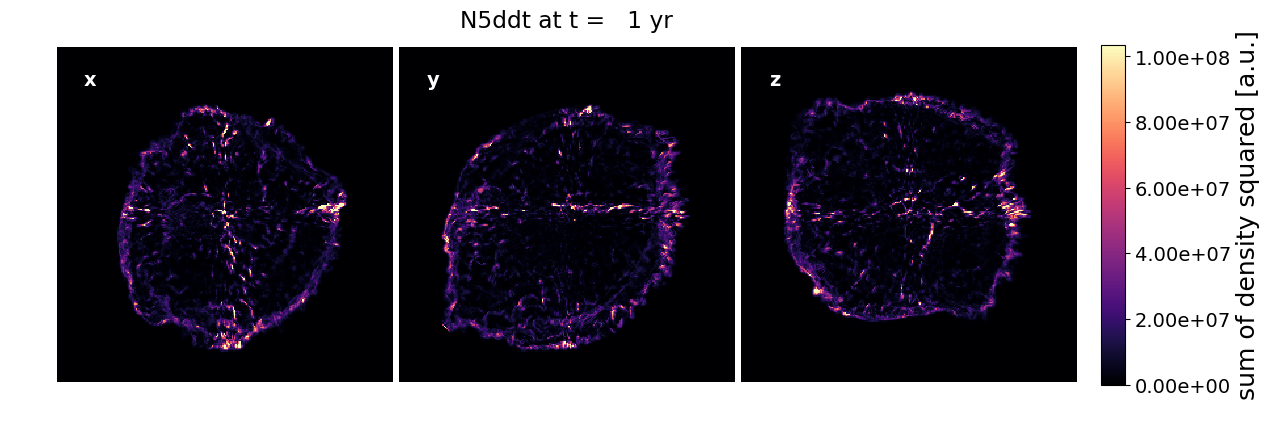}
\includegraphics[width=\widthmap]{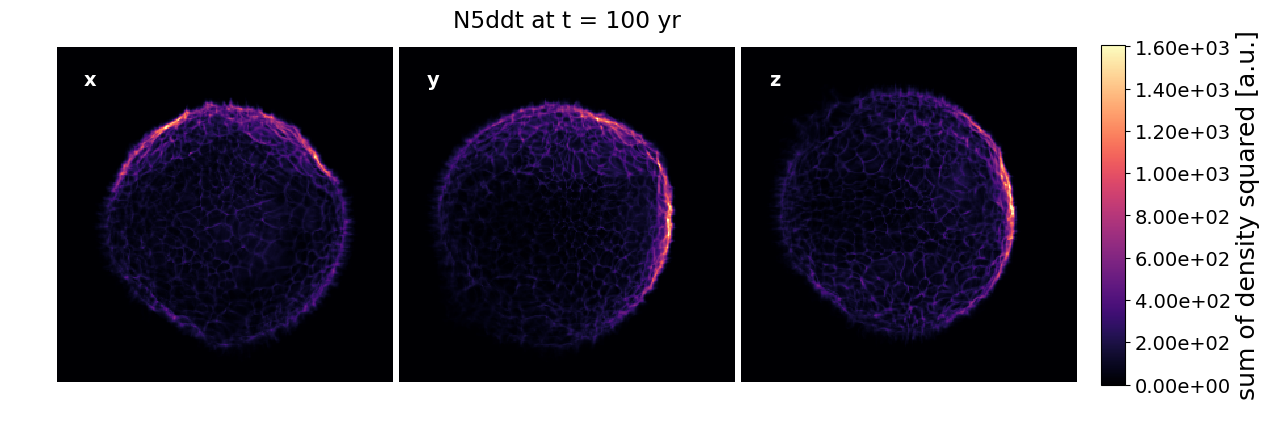}
\includegraphics[width=\widthmap]{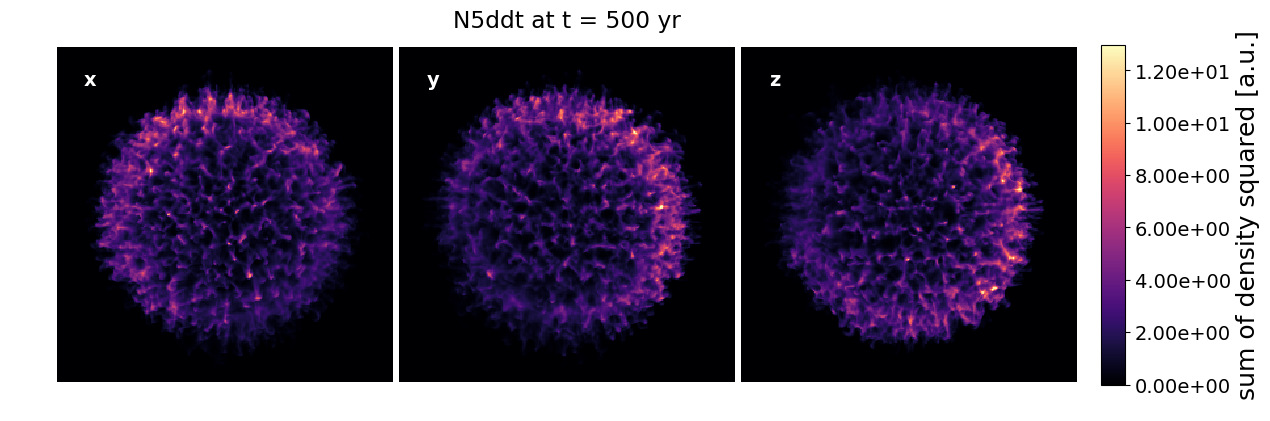}
\caption{Projected maps of density squared for N5ddt. The size of the box is $0.0704$~pc at 1~yr, $4.32$~pc at 100~yr, $13.6$~pc at 500~yr. An animation over time from 1 to 500~yr in steps of 1~yr is available (movie duration: 10~s).
\label{fig:map-prj-N5ddt}}
\end{figure}

\begin{figure}[p]
\includegraphics[width=\widthmap]{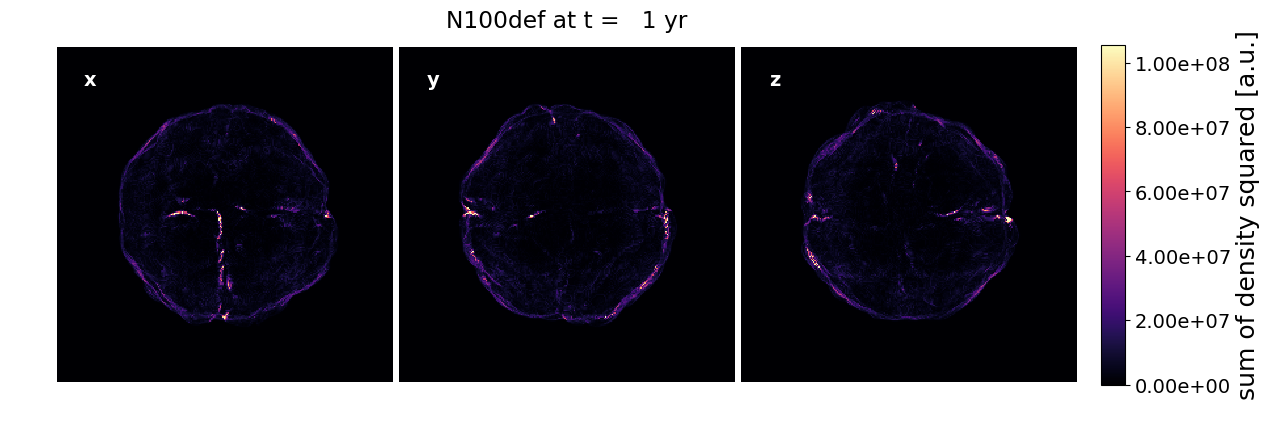}
\includegraphics[width=\widthmap]{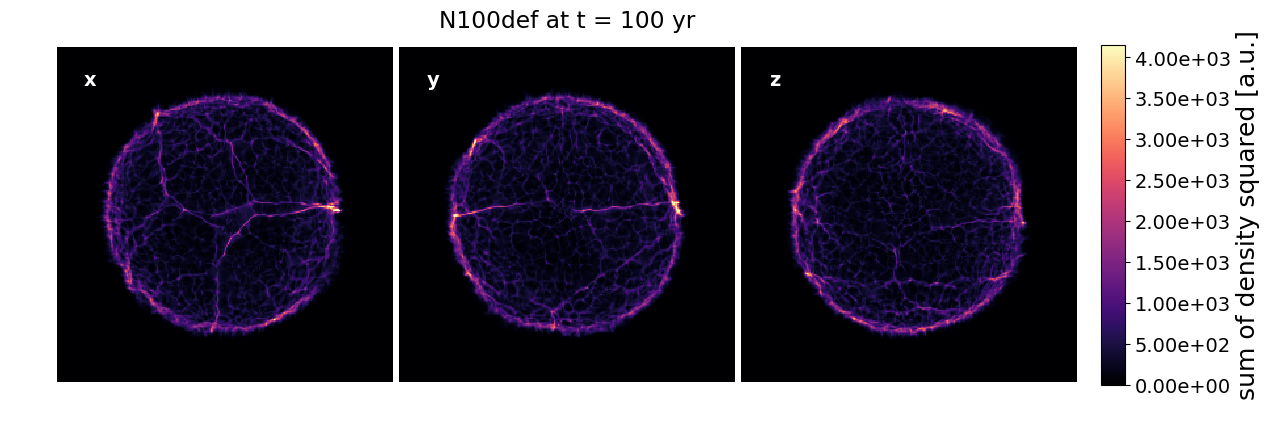}
\includegraphics[width=\widthmap]{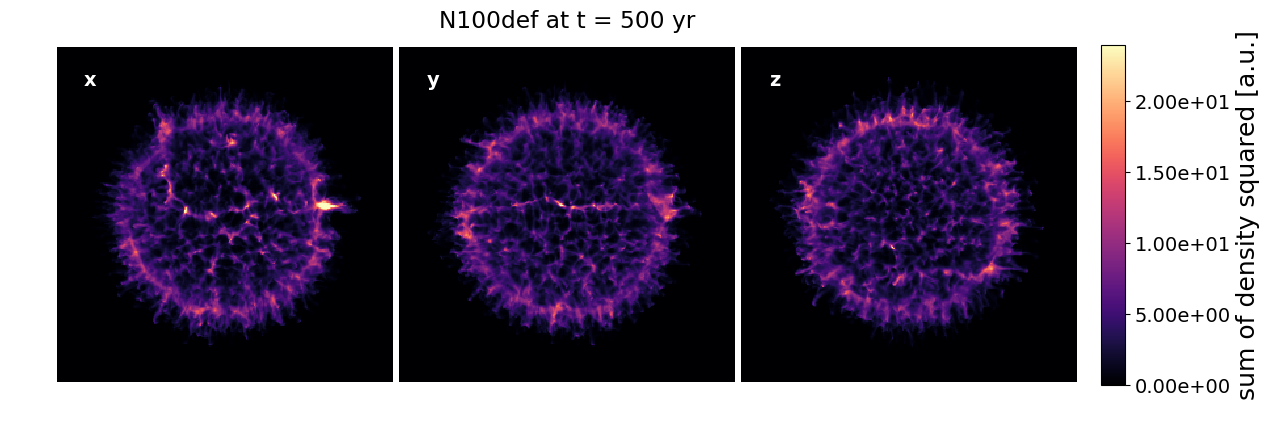}
\caption{Projected maps of density squared for N100def. The size of the box is $0.0438$~pc at 1~yr, $3.20$~pc at 100~yr, $11.6$~pc at 500~yr. An animation over time from 1 to 500~yr in steps of 1~yr is available (movie duration: 10~s).
\label{fig:map-prj-N100def}}
\end{figure}

\begin{figure}[p]
\includegraphics[width=\widthmap]{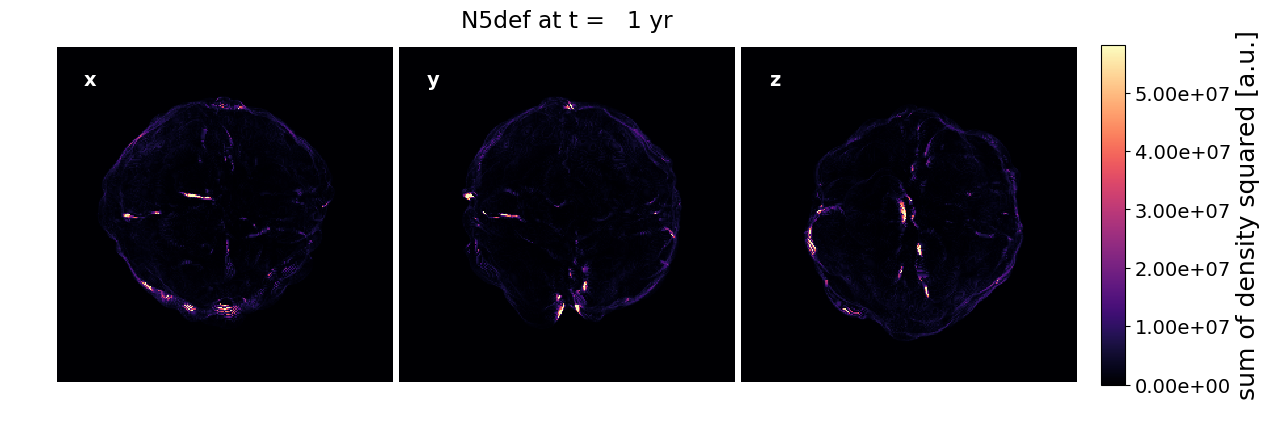}
\includegraphics[width=\widthmap]{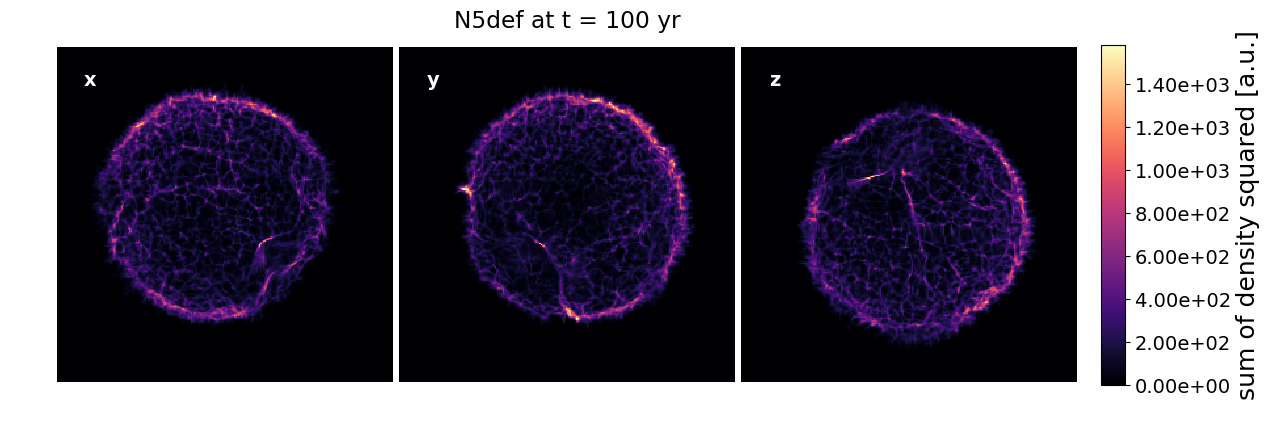}
\includegraphics[width=\widthmap]{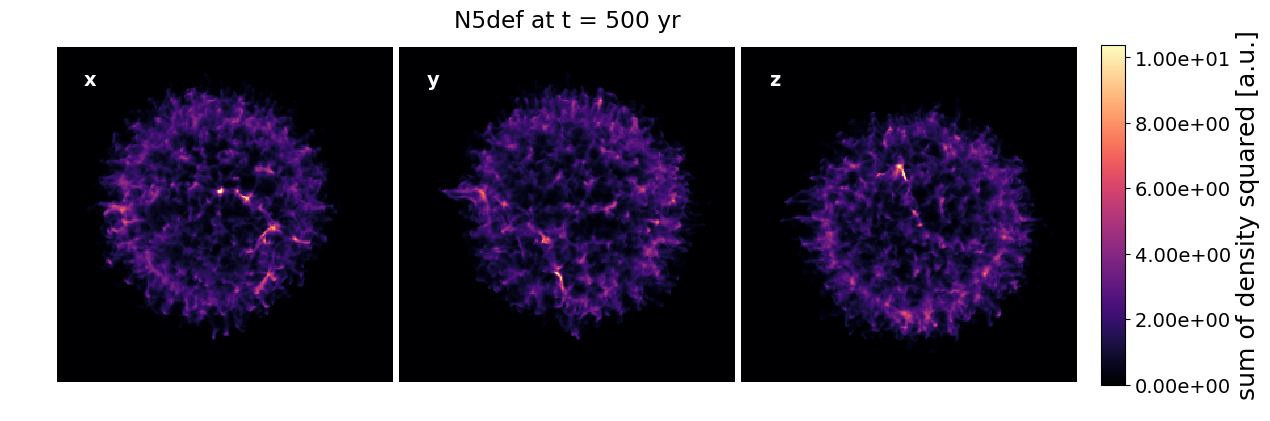}
\caption{Projected maps of density squared for N5def. The size of the box is $0.0384$~pc at 1~yr, $2.82$~pc at 100~yr, $9.36$~pc at 500~yr. An animation over time from 1 to 500~yr in steps of 1~yr is available (movie duration: 10~s).
\label{fig:map-prj-N5def}}
\end{figure}

In Figures~\ref{fig:map-prj-N100ddt}, \ref{fig:map-prj-N5ddt}, \ref{fig:map-prj-N100def}, \ref{fig:map-prj-N5def} we show for each model projections of the density squared in the shocked region, which is a proxy for the (broad-band) thermal emission from the hot plasma (which falls in the X-ray domain, around keV energies). Again we show for each model the map along three orthogonal directions $x$, $y$, $z$, and at three different times 1~yr, 100~yr, 500~yr. Movies from 1~yr to 500~yr in steps of 1~yr are also available. 
Note that, since the emission from the plasma depends on its electronic temperature and ionization state, which are time-dependent, a more precise calculation is needed to produce realistic X-ray maps, which we defer to a future study. These maps allow us to assess, at a glance, what kind of morphological differences can be expected on observations of SNRs from the four SN models. 

For N100ddt, as in Paper~I, maps at 500~yr are rather symmetric, although not as much as simulations made from smooth initial conditions, and so somewhat depend on the observation direction. The edge of the ejecta generally looks brighter, as expected in projection (limb brightening effect). 
In comparison, the N5ddt model displays a distinctive two-sided structure, with one half of the SNR significantly brighter than the other (of course, the effect would not be visible if looking right along the dipolar axis). This is a direct consequence from the varying thickness of the shocked region.

The def models, N100def and N5def, have a more angularly-symmetric shape in projection (although they may not look exactly the same under every direction). Contrary to the ddt case, differences between the N100 and N5 versions are not as obvious. Except for the fact that the N5def SNR is significantly off-centred, but that would not be inferred from the map alone without additional information (like measurements of the radial motion of the ejecta, pointing to the explosion centre). A~distinctive feature of the def models, present for both N100 and N5 versions, is the appearance of bright lines that cross the SNR surface. We will see them in another form in the next section. Finally, we note that these maps in projection, because they show only the matter bounded by the RS and FS, cannot feature the central over-density of the def models (otherwise this would be the dominant feature). 

\subsection{Analysis of the angular structure}
\label{sec:results-healpix}

\paragraph{Edge of the ejecta.}

In Figures~\ref{fig:healpix-N100ddt-CD}, \ref{fig:healpix-N5ddt-CD}, \ref{fig:healpix-N100def-CD}, \ref{fig:healpix-N5def-CD}, on the left panels we show in a Mollweide projection the surface of the contact discontinuity (edge of the ejecta). The function plotted is the relative variations in radius from the explosion centre, $R(\theta,\phi)=(r_{\rm CD}-\langle r_{\rm CD}\rangle)/\langle r_{\rm CD}\rangle$. Regions in tones of red are ahead of the average ($R>0$), while regions in tones of blue are lagging ($R<0$). Compared with the maps of the previous section, these maps contain information in all directions at once. 
On the right panels we show the power spectrum resulting from the expansion in spherical harmonics of the function $R(\theta,\phi)$, as a function of angular wavenumber~$\ell$. 
The projections allow us to quickly visually assess the shape of the ejecta, while the spectra allow us to quantify the asymmetries. 
The power $C_{\ell}$ plotted is normalized in such a way that each grayed bin is the contribution of wavenumber~$\ell$ to the total variance of $R(\theta,\phi)$.
As before the plots are shown at three different times 1~yr, 100~yr, 500~yr. Movies from 1~yr to 500~yr in steps of 1~yr are also available.

\def\widthhealpix{1\textwidth}

\begin{figure}[p]
\centering
\includegraphics[width=\widthhealpix]{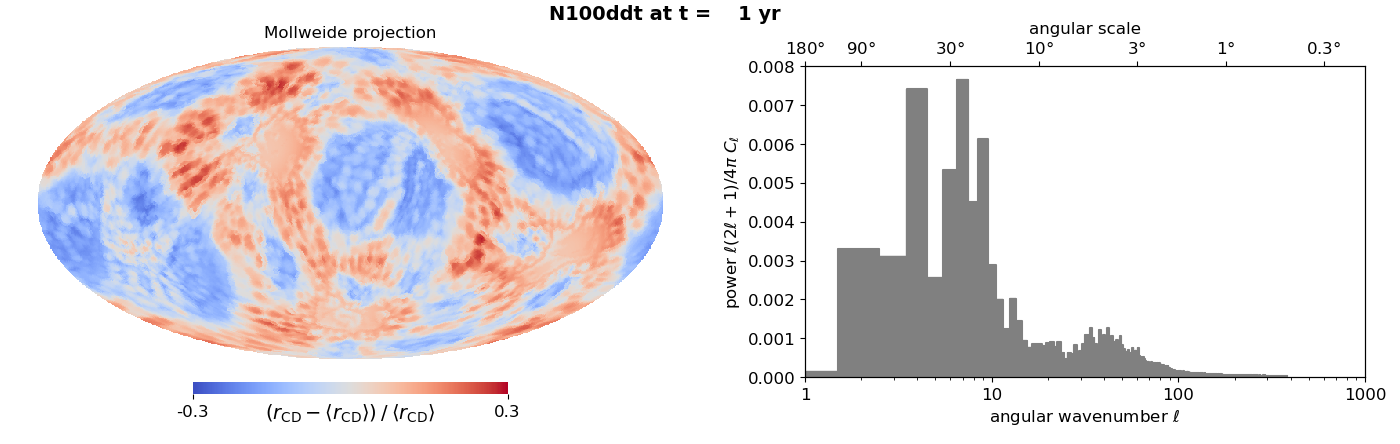}
\includegraphics[width=\widthhealpix]{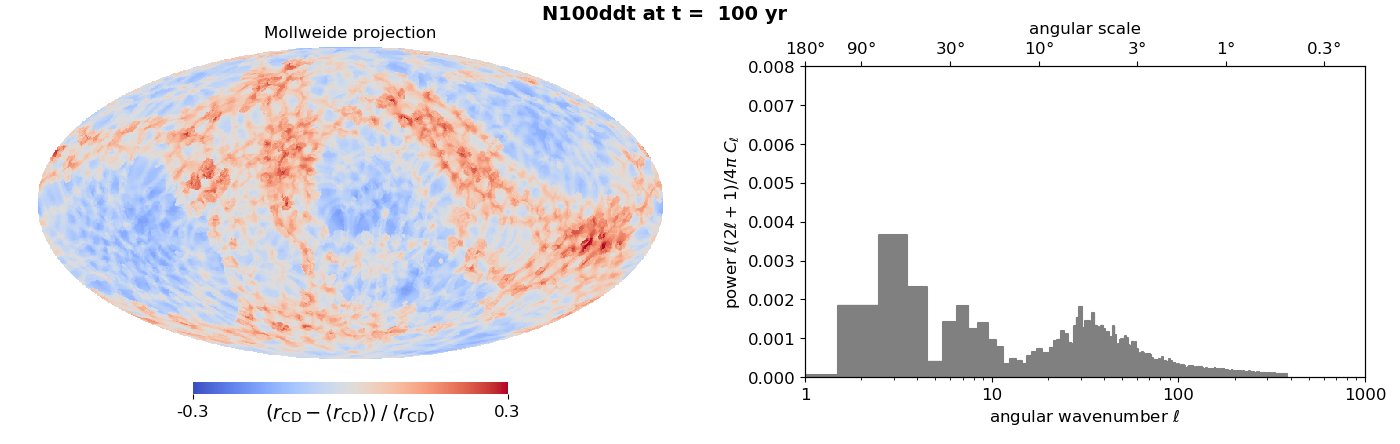}
\includegraphics[width=\widthhealpix]{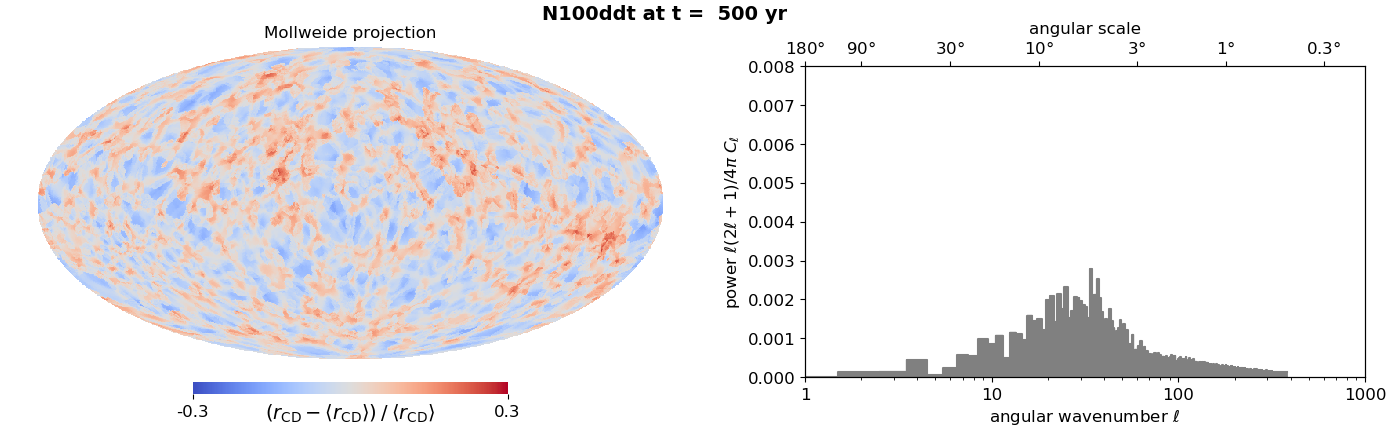}
\caption{Morphology of the contact discontinuity for N100ddt. 
Maps on the left are spherical projections of the radial variations of the location of the wave. We use the Mollweide projection, for all times it is centred on the dipole component of this model at the initial time. Spectra on the right result from an expansion in spherical harmonics of these variations. At angular wavenumber~$\ell$, the typical angular scale probed is $\pi/\ell$, and the power $C_{\ell}$ plotted is normalized in such a way that each grayed bin is the contribution of wavenumber~$\ell$ to the total variance of the radial fluctuations.
As for the previous maps, three times are shown: 1~yr, 100~yr, and 500~yr. An animation over time from 1 to 500~yr in steps of 1~yr is available (movie duration: 10~s).
\label{fig:healpix-N100ddt-CD}}
\end{figure}

\begin{figure}[p]
\centering
\includegraphics[width=\widthhealpix]{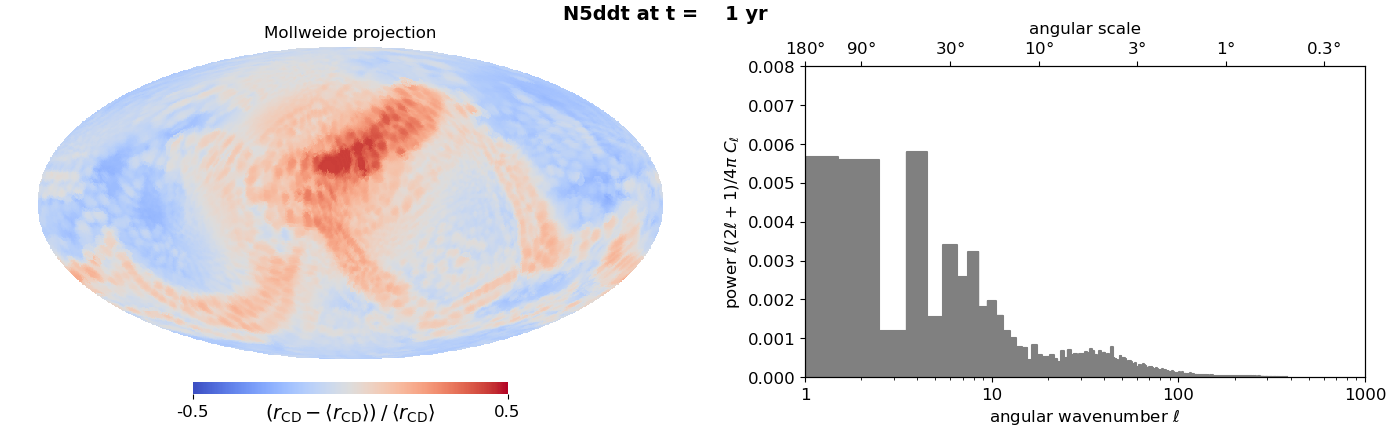}
\includegraphics[width=\widthhealpix]{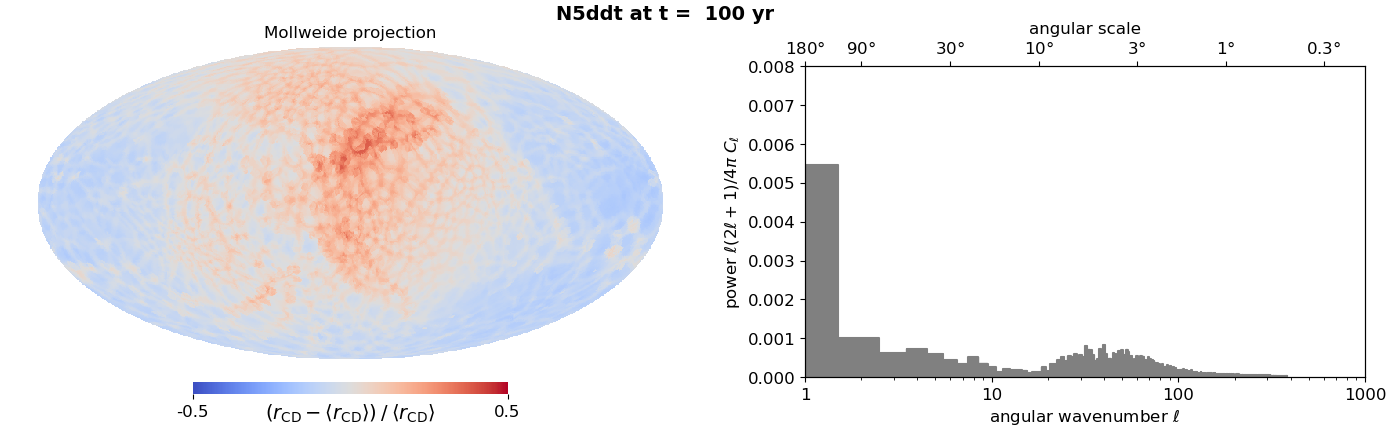}
\includegraphics[width=\widthhealpix]{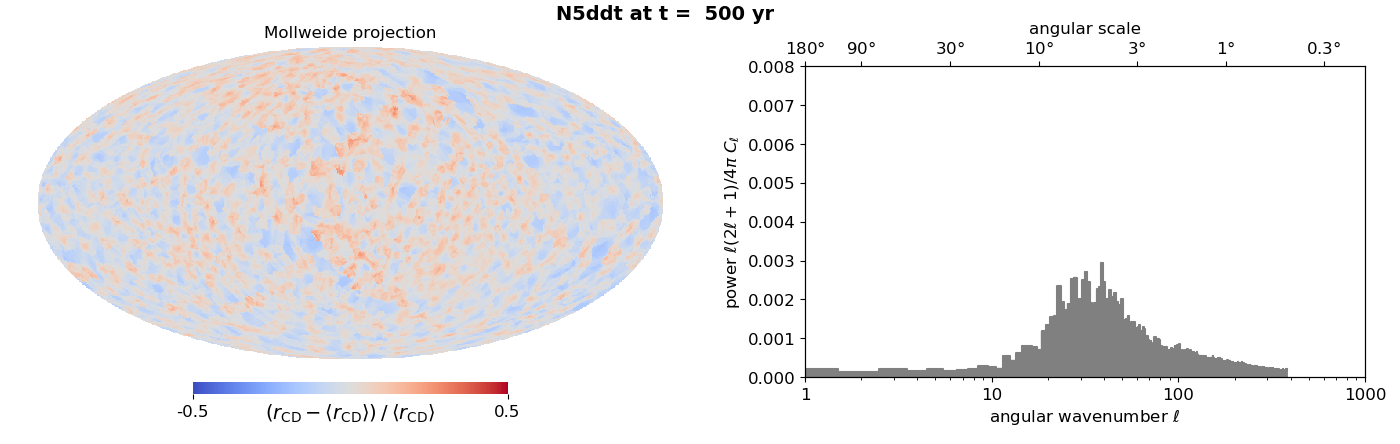}
\caption{Morphology of the contact discontinuity for N5ddt. An animation over time from 1 to 500~yr in steps of 1~yr is available (movie duration: 10~s).
\label{fig:healpix-N5ddt-CD}}
\end{figure}

\begin{figure}[p]
\centering
\includegraphics[width=\widthhealpix]{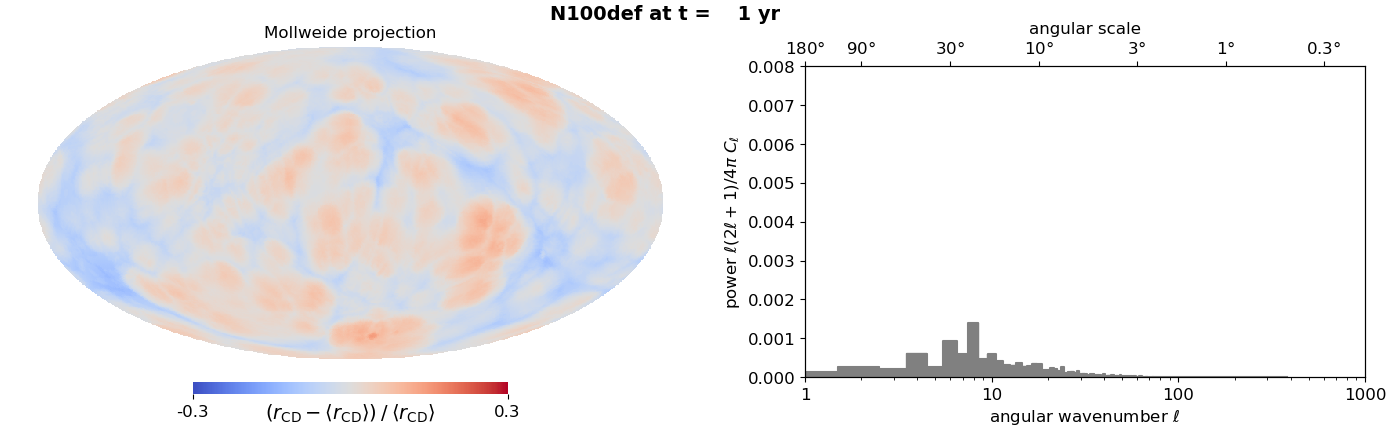}
\includegraphics[width=\widthhealpix]{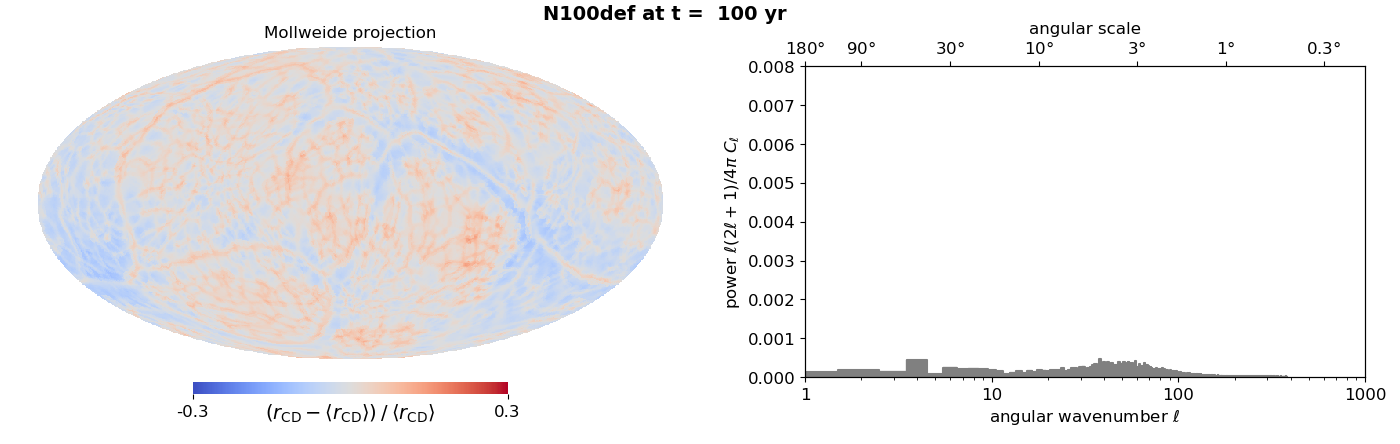}
\includegraphics[width=\widthhealpix]{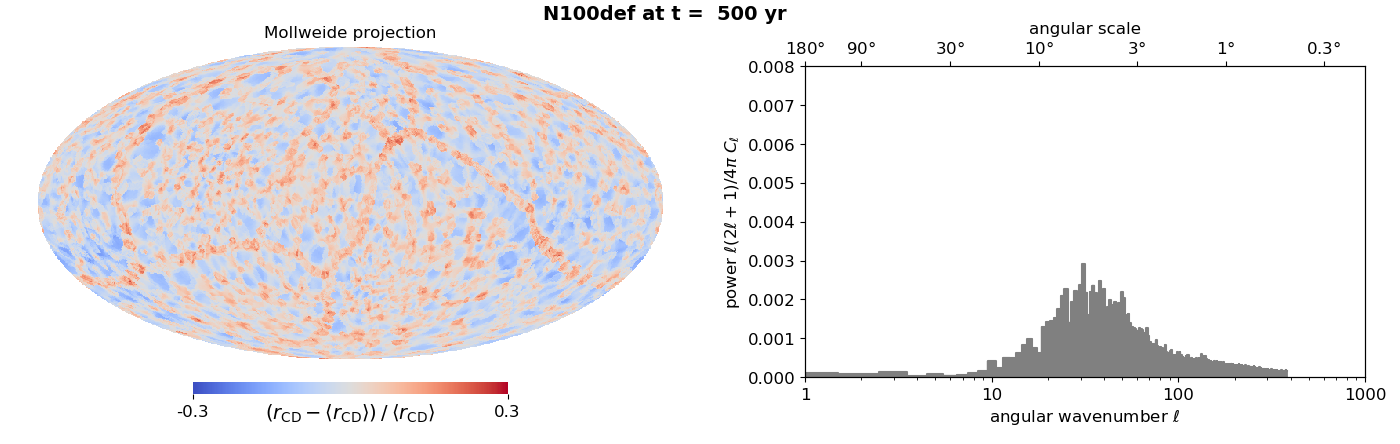}
\caption{Morphology of the contact discontinuity for N100def. An animation over time from 1 to 500~yr in steps of 1~yr is available (movie duration: 10~s).
\label{fig:healpix-N100def-CD}}
\end{figure}

\begin{figure}[p]
\centering
\includegraphics[width=\widthhealpix]{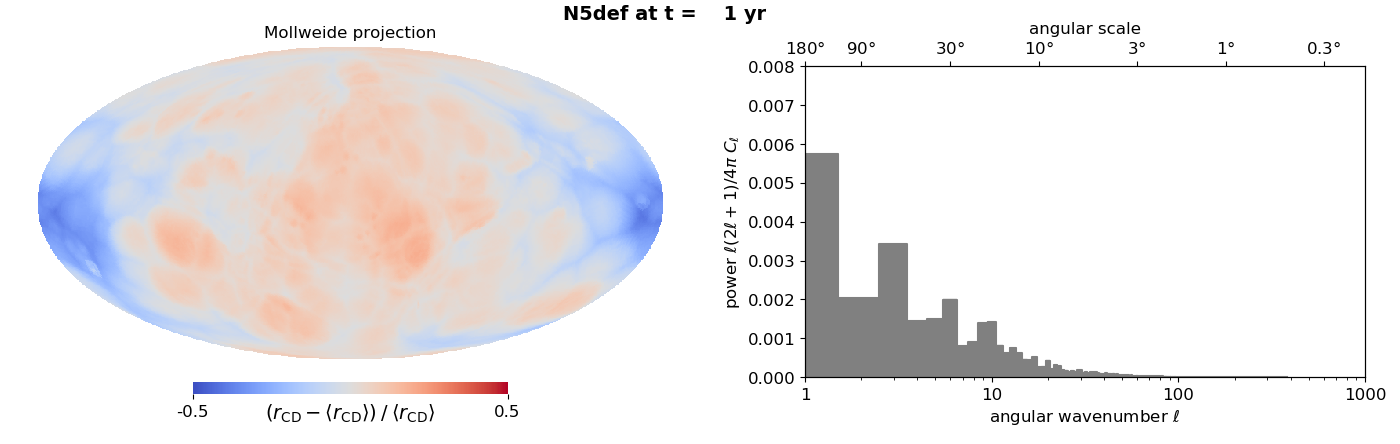}
\includegraphics[width=\widthhealpix]{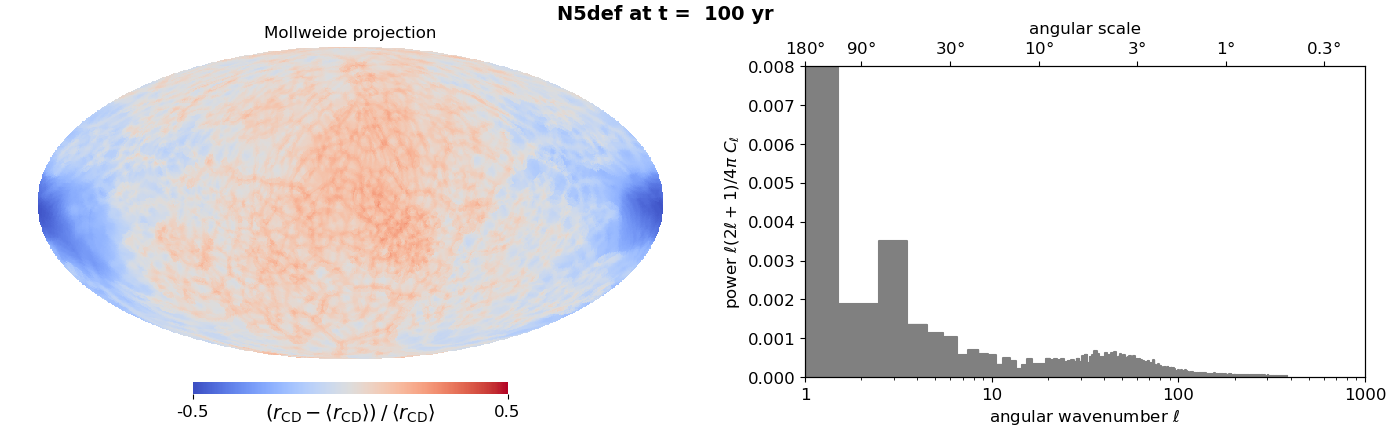}
\includegraphics[width=\widthhealpix]{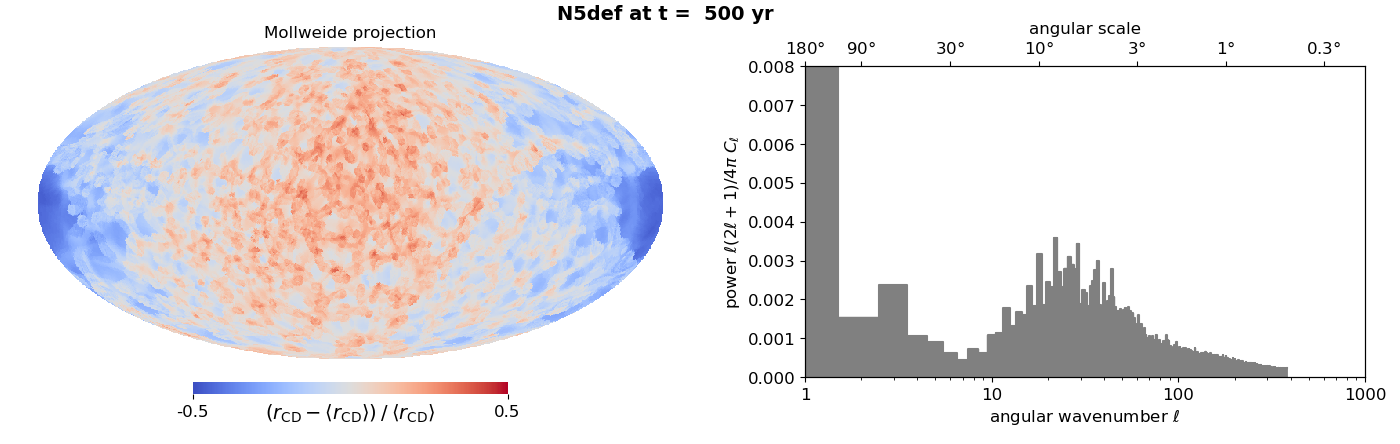}
\caption{Morphology of the contact discontinuity for N5def. An animation over time from 1 to 500~yr in steps of 1~yr is available (movie duration: 10~s).
\label{fig:healpix-N5def-CD}}
\end{figure}

The CD surface at 1~yr highlights the morphological differences between the different SN models. First we see the difference between N100 models, for which power peaks at intermediate angular scales, and N5 models, for which power peaks at the dipole component $\ell=1$. For N5ddt the dipole takes the form of a protrusion: a relatively small area is ahead of average by 50\%, while a larger area is lagging by half this amount. For N5def the dipole is more balanced, with roughly two opposite halves. For N5ddt the dipole strength decreases over time via the mechanism that we have discussed previously, the ejecta become more regular over time, although an asymmetry is still visible at 500~yr. For N5def the dipole remains at all times, as the SNR is effectively off-centred as we have seen. 
Second we see differences between ddt and def models, with the latter having weaker asymmetries (note that the scale for the projection maps has been set to the same value for the ddt and def versions of a given ignition setup).

On top of these large scale effects, that stem from the chosen SN model, we see the RTI steadily growing over time, at smaller scales. After a couple hundred years it competes with the initial conditions in shaping the remnant, and at 500~yr it controls most of what is immediately visible by eye.

Another effect, a~distinctive feature of the def models, that was not expected, is the progressive appearance over time of contiguous lines, consistently ahead of average. They are particularly clear on the maps for the N100def model (Figure~\ref{fig:healpix-N100def-CD}), even though they contribute little power. These protruding structures match the lines previously reported on the projected density maps (Figures~\ref{fig:map-prj-N100def} and~\ref{fig:map-prj-N5def}). They exhibit an irregular pattern that does not match the geometry of our simulation mesh. Looking at the time evolution, we see that some of them grow on top of pre-existing ejecta ridges, while some of them grow at the bottom of ejecta trenches. They have dynamics completely separate from that of the RTI. After a few hundred years they form a network of seam lines on the ejecta surface, separating the ejecta into patches. 

\paragraph{Width of the shocked region.}

In this paper we are not showing the individual plots for the RS and FS, which show less variations than for the CD (these plots are present in the online repository). As shown in Paper~I, the RS, being always close from the CD for such young SNRs, shows some echo of the CD fluctuations (feet of the RTI). The FS initially delineates the ejecta surface, and over time detaches from it and gets more regular, although it may still show some imprint of the turbulent interior (fingers of the RTI). These effects are visible for each of the SN models, to varying extent. 

\begin{figure}[p]
\centering
\includegraphics[width=\widthhealpix]{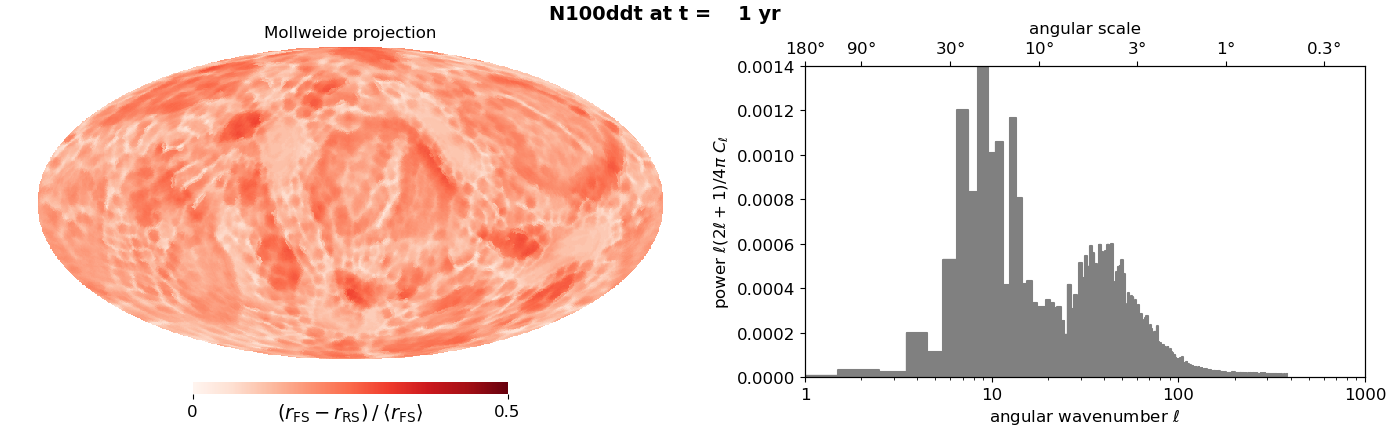}
\includegraphics[width=\widthhealpix]{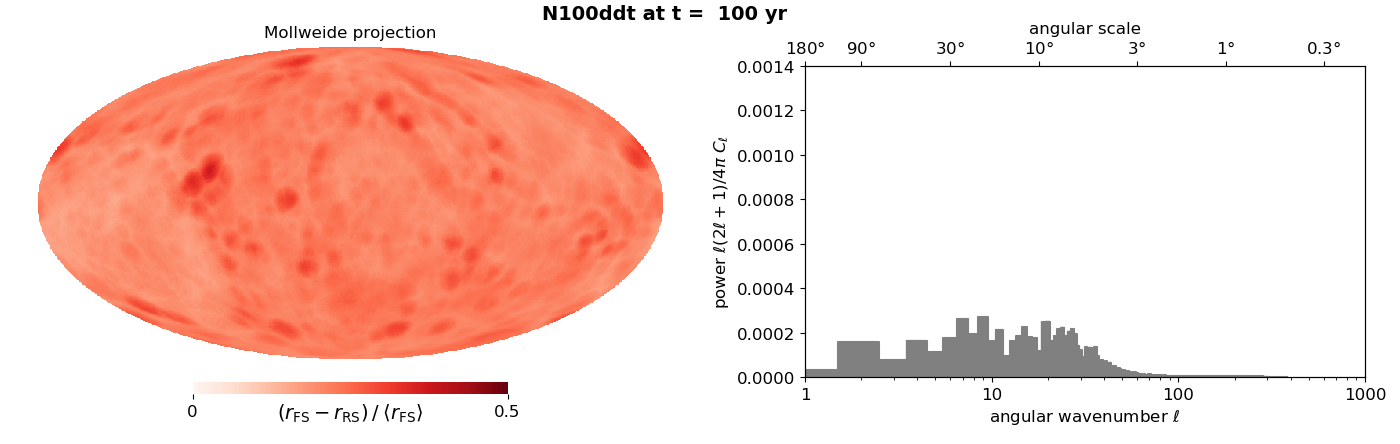}
\includegraphics[width=\widthhealpix]{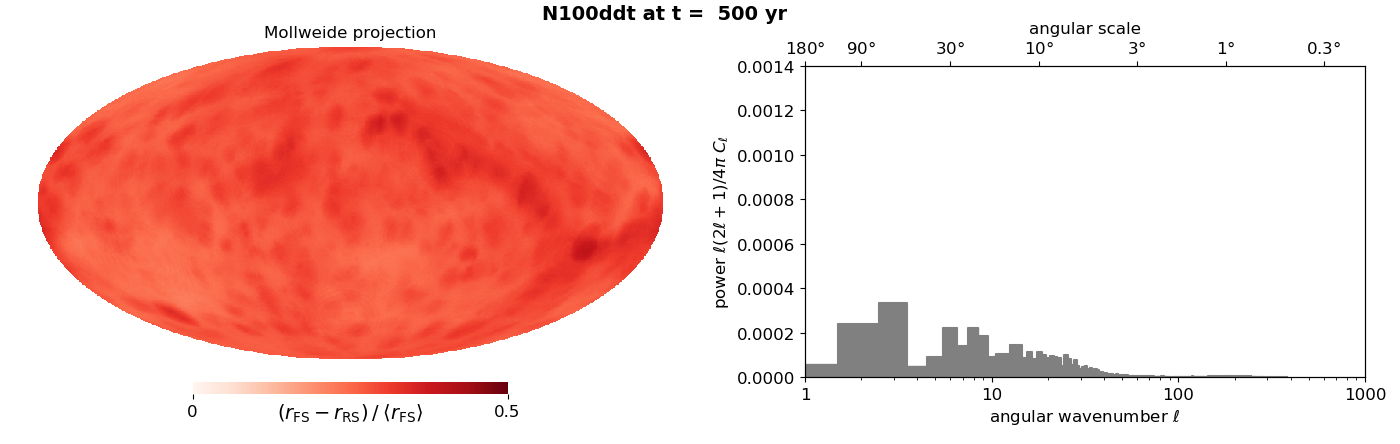}
\caption{Width of the shocked region for N100ddt. 
Maps on the left are spherical projections of the relative width of the shocked region. Spectra on the right result from an expansion in spherical harmonics of these functions. Details are the same as for the plots of the CD.
Three times are shown: 1~yr, 100~yr, and 500~yr. An animation over time from 1 to 500~yr in steps of 1~yr is available (movie duration: 10~s).
\label{fig:healpix-N100ddt-FS-RS}}
\end{figure}

\begin{figure}[p]
\centering
\includegraphics[width=\widthhealpix]{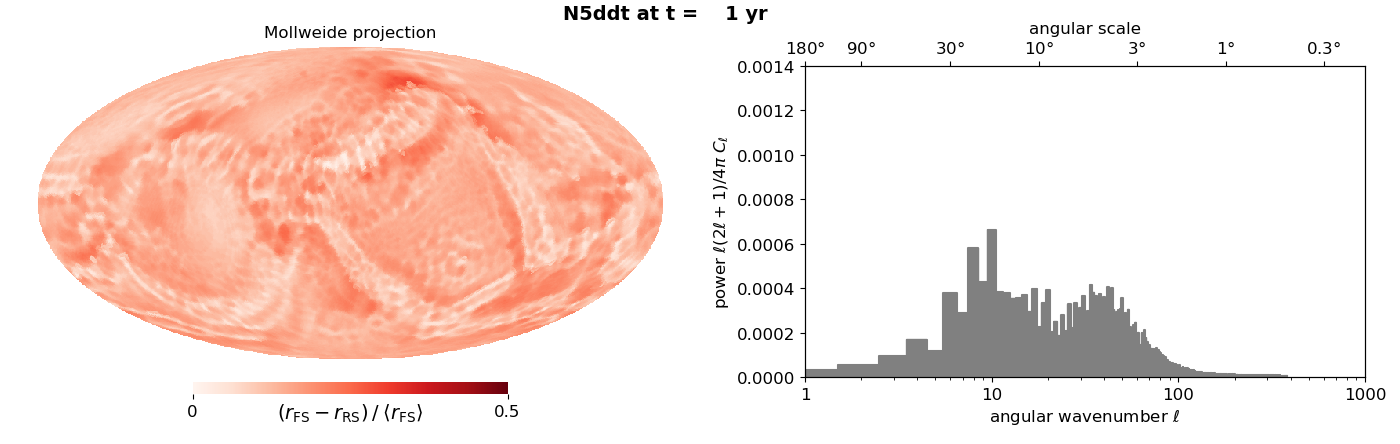}
\includegraphics[width=\widthhealpix]{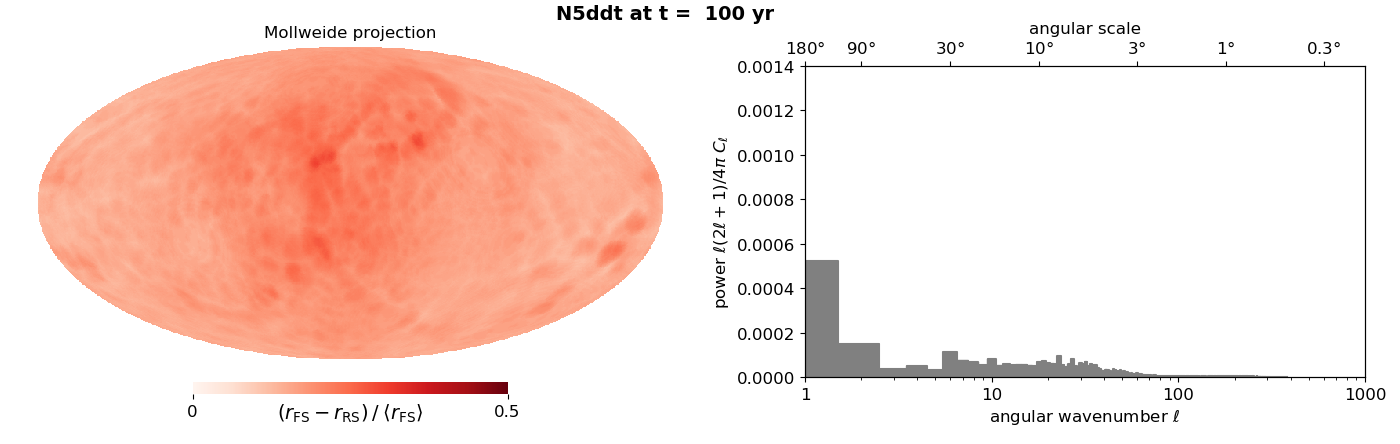}
\includegraphics[width=\widthhealpix]{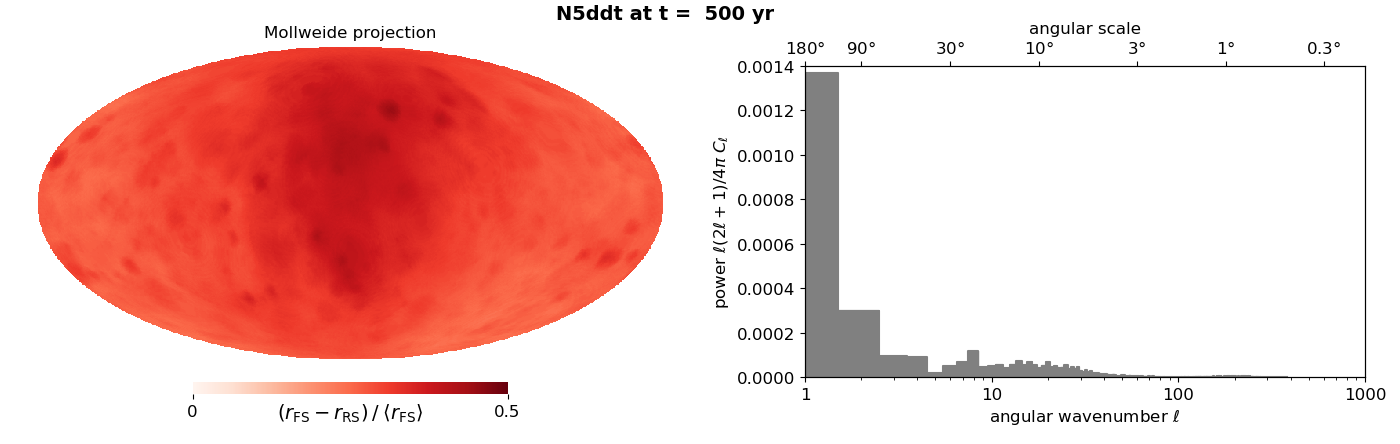}
\caption{Width of the shocked region for N5ddt. An animation over time from 1 to 500~yr in steps of 1~yr is available (movie duration: 10~s).
\label{fig:healpix-N5ddt-FS-RS}}
\end{figure}

\begin{figure}[p]
\centering
\includegraphics[width=\widthhealpix]{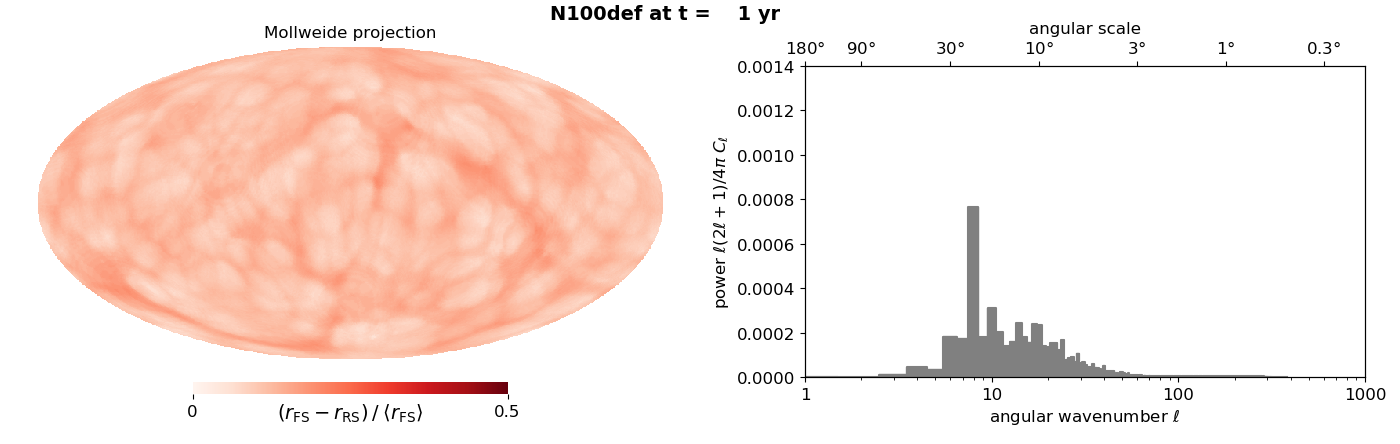}
\includegraphics[width=\widthhealpix]{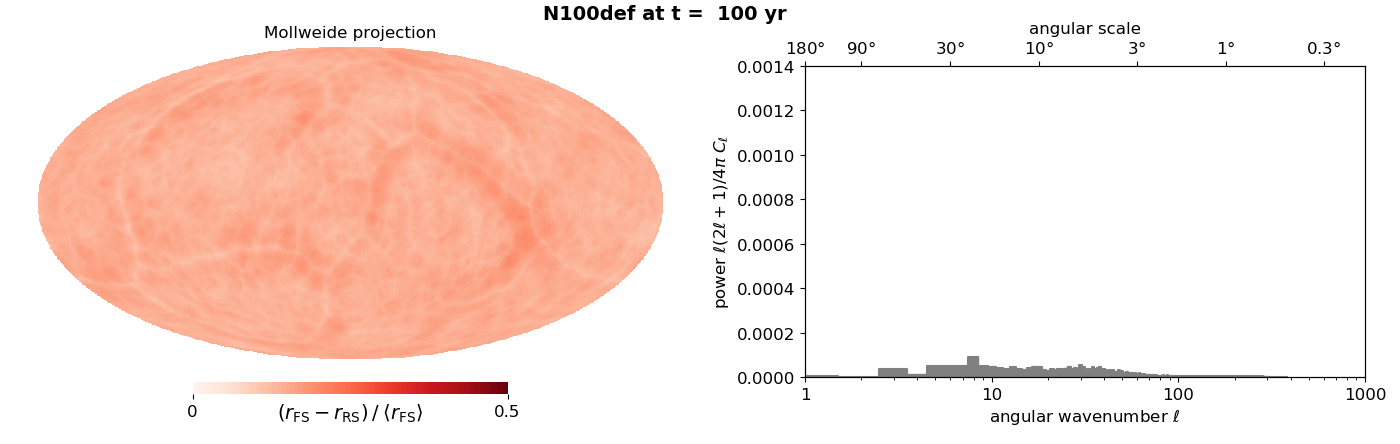}
\includegraphics[width=\widthhealpix]{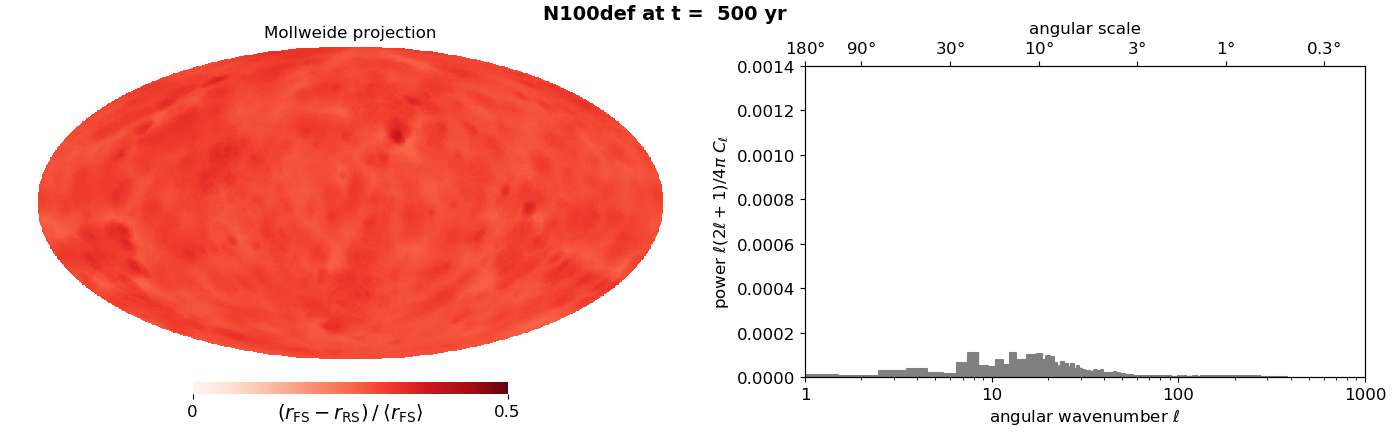}
\caption{Width of the shocked region for N100def. An animation over time from 1 to 500~yr in steps of 1~yr is available (movie duration: 10~s).
\label{fig:healpix-N100def-FS-RS}}
\end{figure}

\begin{figure}[p]
\centering
\includegraphics[width=\widthhealpix]{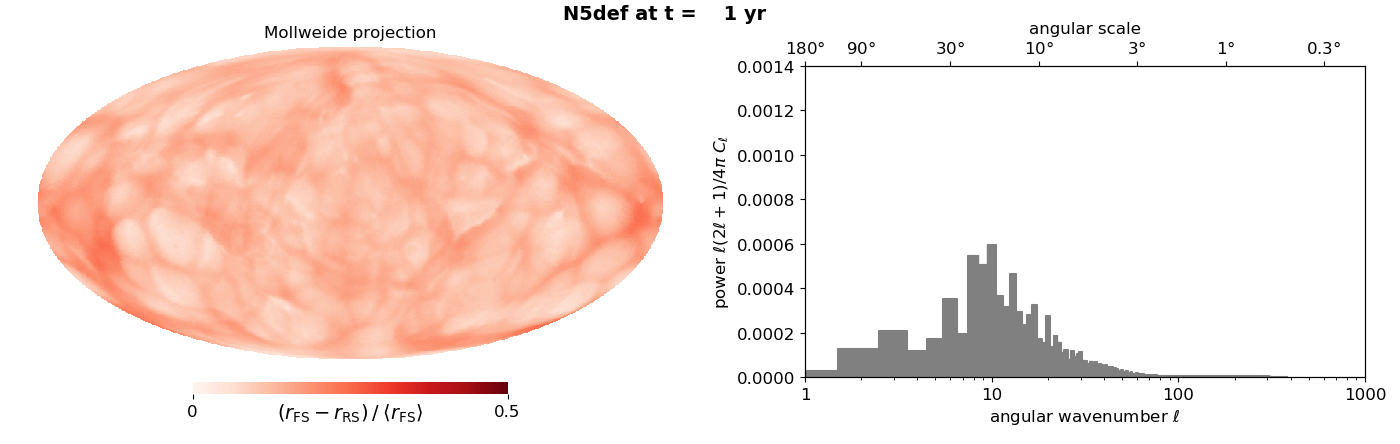}
\includegraphics[width=\widthhealpix]{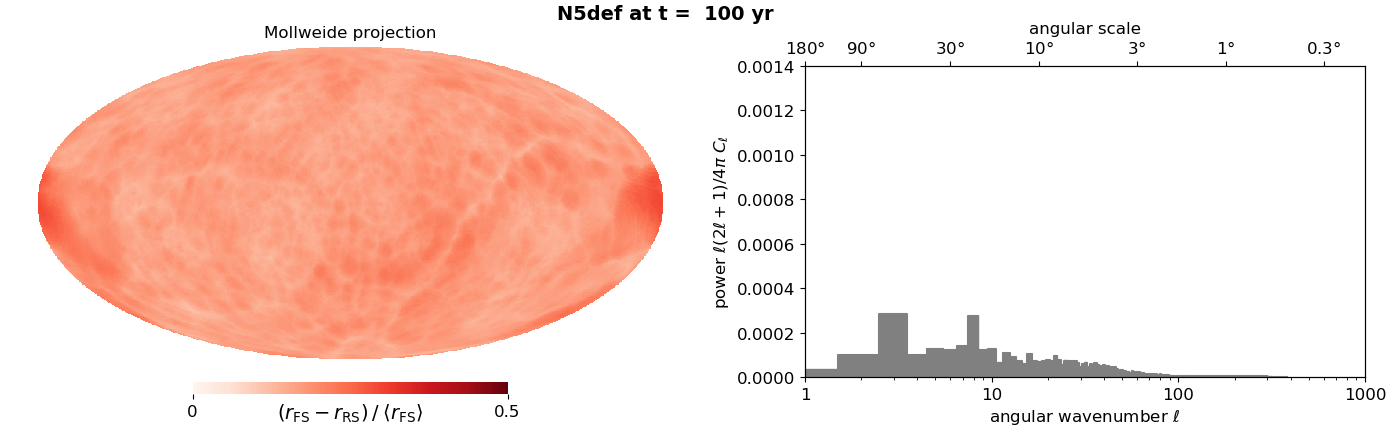}
\includegraphics[width=\widthhealpix]{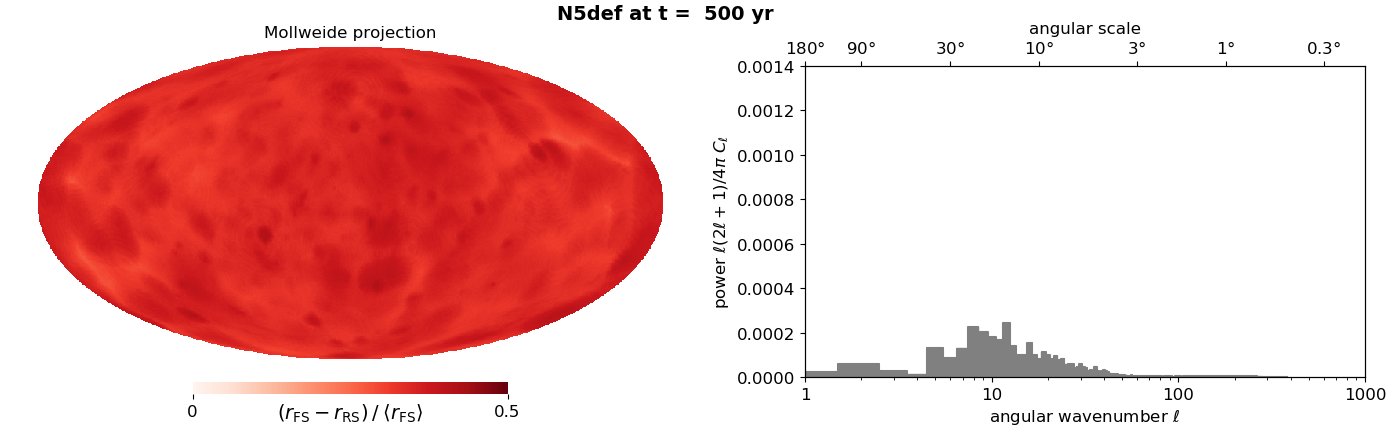}
\caption{Width of the shocked region for N5def. An animation over time from 1 to 500~yr in steps of 1~yr is available (movie duration: 10~s).
\label{fig:healpix-N5def-FS-RS}}
\end{figure}

Instead, in Figures~\ref{fig:healpix-N100ddt-FS-RS}, \ref{fig:healpix-N5ddt-FS-RS}, \ref{fig:healpix-N100def-FS-RS}, \ref{fig:healpix-N5def-FS-RS} we are showing the width of the shocked region, that is the radial distance between the RS and the FS, normalized by the average size of the SNR, taken to be average radius of the FS. So explicitly we consider the function $R(\theta,\phi)=(r_{\rm FS}- r_{\rm RS})/\langle r_{\rm FS}\rangle$. We now always have $R>0$, and we are using tints of red on the map. For each models, we see the growth of the shocked region over time. Initially, it naturally bears the imprint of the shape of the ejecta's edge shown above, from which it emerges. As the shocks travel away, and their surface progressively regularizes through the effect of the high pressure, the shocked region becomes globally more homogeneous in time -- except for N5ddt. For all models, small scale variations visible at 500~yr are mostly caused by a few strongly growing RT fingers that are pushing the FS. At this age, the shocked region makes about 20\% to 50\% of the SNR radial size, with the thickest models being N5ddt (on one side) and N5def (on most of its surface).

In line with what we have seen before on slices, N5ddt has a clearly bipolar morphology, with the shocked region about twice as large on one side than on the other. For the N5 models, comparing the plots for the CD (Figures~\ref{fig:healpix-N5ddt-CD} and~\ref{fig:healpix-N5def-CD}) and for FS-RS (Figures~\ref{fig:healpix-N5ddt-FS-RS} and~\ref{fig:healpix-N5def-FS-RS}) illuminates that N5ddt and N5def are asymmetric in completely different ways. For N5ddt, the ejecta surface is initially dipolar, but regularizes over time, while the shocked region follows the opposite trend. As explained before, this stems from the fact that the most extended ejecta are also the most diluted, and so the most easily decelerated ones. This ends up being a two-sided SNR. For N5def, the ejecta are skewed on one side at all times, but the shocked region quickly gets pretty uniform. This makes a regular-looking but offset SNR.

Finally, for the def models, we note that the seams noticed on the ejecta surface are also present. They are particularly visible around 100~yr. They are mostly caused by the fact that the RS is at larger radii than average in these regions -- meaning that it is travelling inward slower than average. So the picture that emerges is that these structures are caused by ejecta that, from the beginning, have greater momentum than average: these ejecta move faster, hence the protruding lines on the ejecta surface (Figures~\ref{fig:healpix-N100def-CD} and~\ref{fig:healpix-N5def-CD}), and thus generate a weaker RS behind them, hence the narrow lines on the shocked region width (Figures~\ref{fig:healpix-N100def-FS-RS} and~\ref{fig:healpix-N5def-FS-RS}). Since the initial SN profiles are homologous, differences in momentum at a given radius come from differences in density, indeed the seams show up on the maps of the projected density squared (Figures~\ref{fig:map-prj-N100def} and~\ref{fig:map-prj-N5def}). For N100def, a~closer inspection of the initial mass density at 100~s reveals that a network of over-dense ridges is indeed present in the outer ejecta layers, below the ejecta surface. A~3D visualization of this structure is available online.\footnote{Hosted on Sketchfab at \url{https://skfb.ly/6VDv7}.}  Its origin appears to be connected with the dynamics of the deflagration fronts during the SN phase, previously mentioned and shown in the Appendix. We interpret the network of over-dense ejecta as the outer boundaries between burning regions. It is this structure that eventually pierces through and becomes apparent during the SNR phase.

\paragraph{Angular power as a function of time.}

\begin{figure}[p]
\includegraphics[width=1\textwidth]{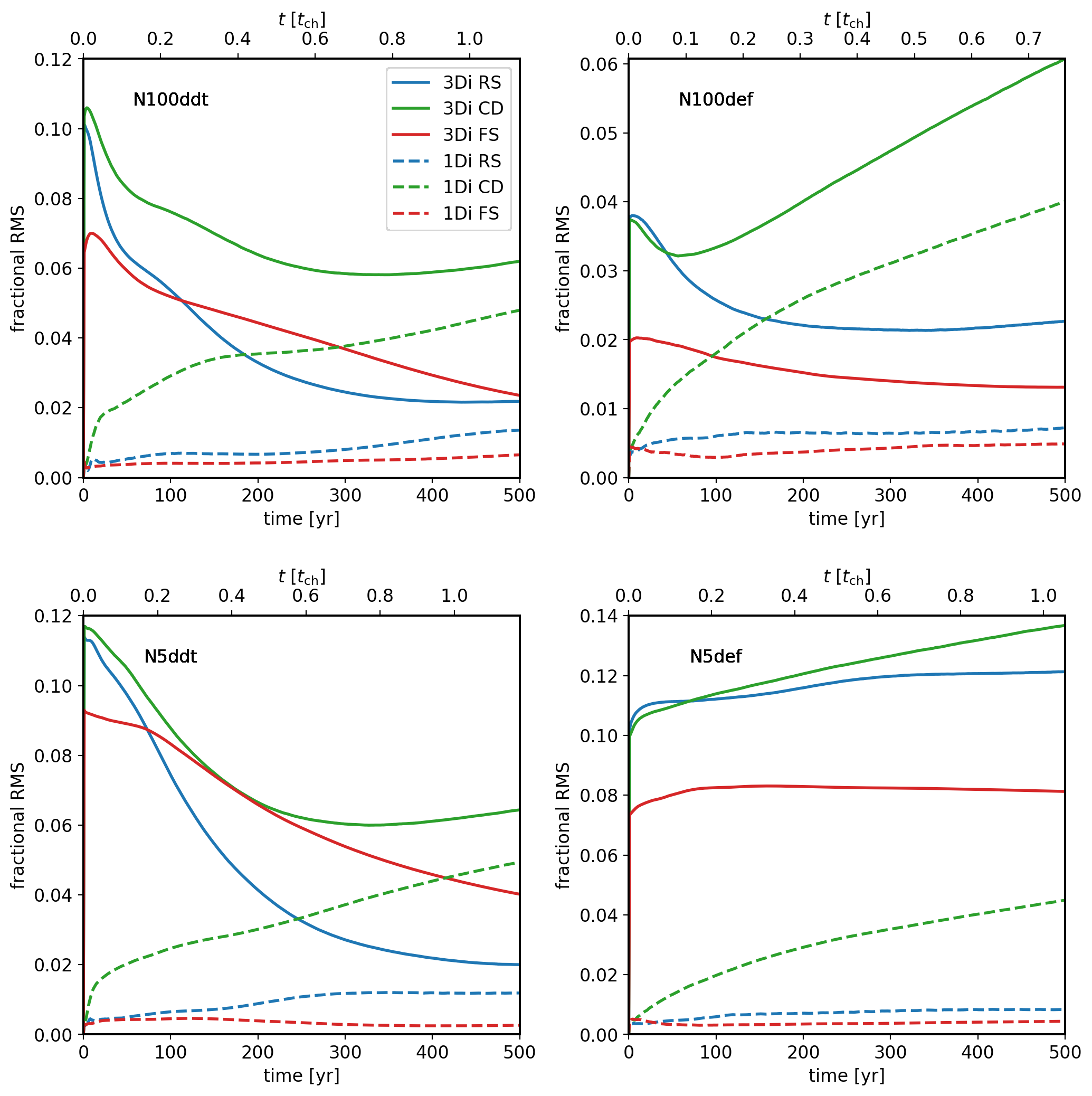}
\caption{Total fractional RMS (square-root of the angular power) as a function of time, for the radial fluctuations of the three wave fronts: FS in red, CD in green, RS in blue. 
Time is indicated in years and in characteristic timescale as defined by equation~(\ref{eq:t_ch}). Note that the former is the same for all models, while the latter depends on the model. Two cases are compared for the initial conditions: spherically symmetric ejecta (1Di, dashed curves) versus asymmetric ejecta (3Di, solid lines). 
\label{fig:power}}
\end{figure}

\begin{figure}[p]
\includegraphics[width=1\textwidth]{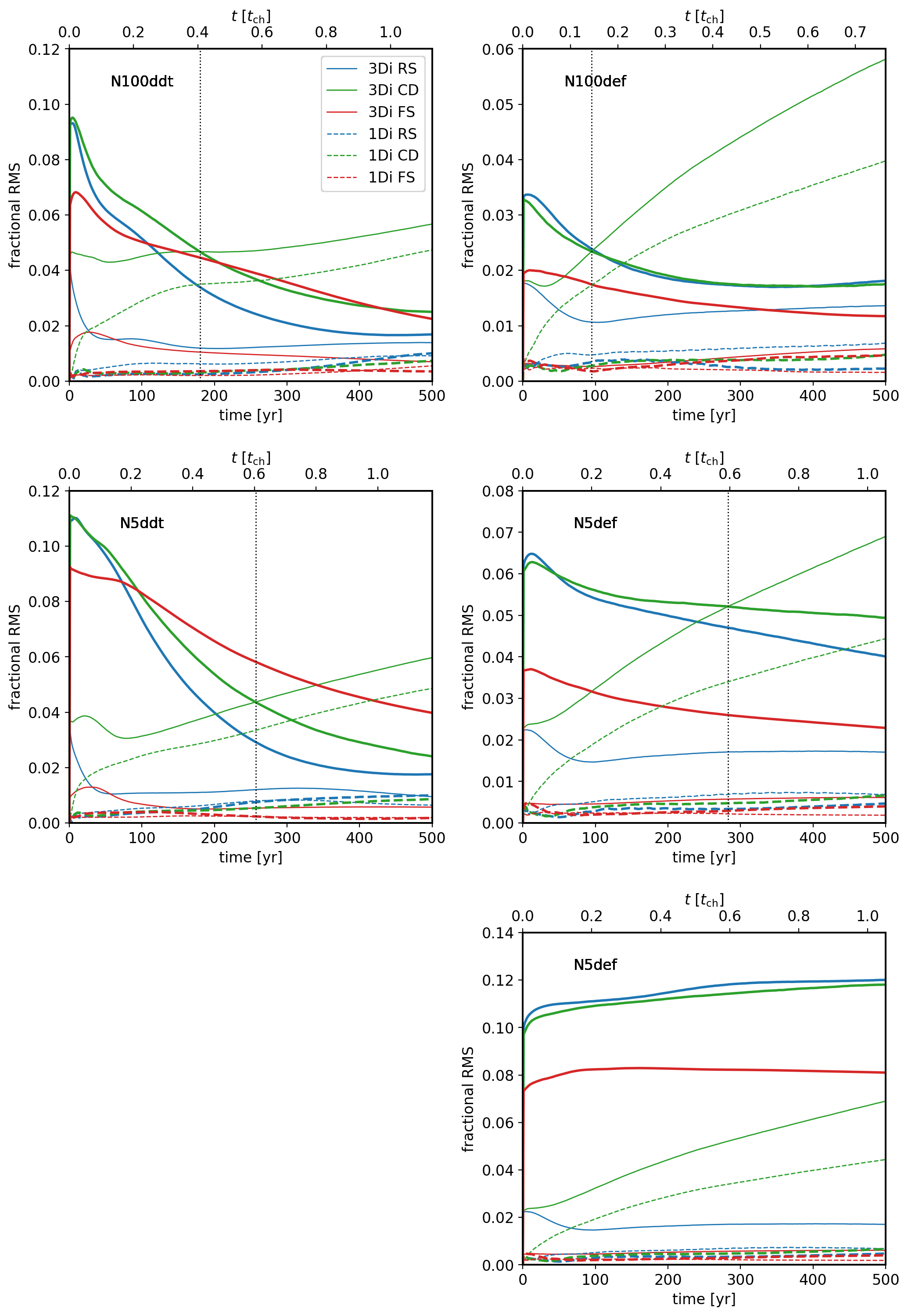}
\caption{Same as figure~\ref{fig:power}, with the fractional RMS plotted separately at large scales ($\ell<10$, thick lines) and at small scales ($\ell>10$, thin lines). The vertical dotted line indicates the time at which the two curves intersect for the CD for the 3Di simulation. For the N5def case the $\ell=1$ contribution has been removed so that the decay of the large scale modes be visible, otherwise the curves never intersect as shown in the bottom-rightmost plot. 
\label{fig:power-split}}
\end{figure}

To finish the presentation of the results, we show the evolution of a global quantity: the summed angular power of the function $R(\theta,\phi)=(r-\langle r\rangle)/\langle r\rangle$. At a given time, the power in a given range of angular scales~$\ell$ is the area of the shaded region on the angular spectra between these scales. Its time evolution tells us the contribution of these scales in the shaping of the SNR morphology. 

In Figure~\ref{fig:power} we plot the total angular power (summed up from $\ell=1$ to $\ell_{\rm max}=383$) as a function of time, for the four SN models.
On each plot we show the power for the three wave fronts: CD in green (corresponding to the spectra of Figures~\ref{fig:healpix-N100ddt-CD} to~\ref{fig:healpix-N5def-CD}), RS in blue, and FS in red. 
On each plot we show the power for the actual SN model (3Di, solid lines) and for the angularly-averaged version of it (1Di, dashed lines). The case of effectively 1D initial conditions only exhibits the effects due purely to the SNR phase, and so as shown in Paper~I comparing the two setups allows us to isolate the contribution from the SN initial conditions. In the 1Di case not much is expected to happen at the RS and FS, while the curve for the CD shows the continued growth of the RTI. In the 3Di case there is power from the initial conditions. For the FS and RS it only decays in time, for the CD it starts decreasing as the memory of the initial conditions fades then eventually increases as the RTI makes its presence felt; the long-term behaviour at the CD is similar in the 1Di and 3Di cases because it is produced by the same RTI. We therefore see two regimes (except for the N5def model, for reasons explained below).

We observed on the power spectra for the 1Di setups that the RTI does not reach below about $\ell=10$ (and peaks around $\ell=50-60$ depending on the models), so we can separate the contributions from the SN phase and the SNR phase. In Figure~\ref{fig:power-split} we show separately the power at small scales (large $\ell$, thin lines) and at large scales (small $\ell$, thick lines). For the RS and FS this makes little difference since most of the power is at the largest scales. For the CD we see the SN modes that decay in time, versus the SNR modes that grow in time. The time at which the curves cross for the CD (for the 3Di case) is marked by a vertical line. This can serve as an estimate of the timescale at which the SN-phase turbulence and the SNR-phase instabilities have similar impact on the morphology of the SNR. At earlier times the SNR is dominated by the SN initial conditions, at later times the SNR is dominated by processes intrinsic to the SNR phase.
The N5def model is a special case. We have seen that it remains off-centred, and so exhibits a dipole component at all times. As a result the total power, as well as the power at small scales $\ell<10$, plateaus or even increases in time (bottom-rightmost plot of Figures~\ref{fig:power} and~\ref{fig:power-split}). So we removed the $\ell=1$ contribution (only for the N5def model), to be able to see the decay of the other SN modes, and apply our timescale finding procedure. 
The transition times are N100ddt: 180~yr, N5ddt: 260~yr, N100def: 95~yr, N5def: 280~yr. We see that the N5 models remain under the SN influence for a longer time than the corresponding N100 models. This is consistent with the N5 ignition setups being more asymmetric than the N100 ignition setups. Also the transition time is twice as large for the ddt than for the def model for the N100 case, but it is similar (and in the opposite direction) for the N5 case. So we cannot readily generalize on the effect of deflagration vs.\ deflagration plus detonation. 
\\~

\section{Discussion}
\label{sec:discussion}

In this section we mention other physical processes that can shape the SNR, and comment on the implications of our findings for SNR observations. 

\subsection{Other physical effects}
\label{sec:discussion-other}

We first discuss how some effects, that were not included in the simulations, may affect the interpretation of our results.

\paragraph{Heating from radioactive decay.}

The most important radioactive species is ${}^{56}\mathrm{Ni}$, which powers the SN light curve.
It decays via beta decay to iron in two steps: 
${}_{28}^{56}\mathrm{Ni} \rightarrow {}_{27}^{56}\mathrm{Co} \rightarrow {}_{26}^{56}\mathrm{Fe}$, 
with half-lives of respectively 6~days and 77~days.
The mass of $^{56}{\rm Ni}$ synthesized in the models is,  
N100ddt: $0.60 M_\odot$ = 54\% $M_{\rm ej}$, 
N5ddt  : $0.97 M_\odot$ = 69\% $M_{\rm ej}$, 
N100def: $0.35 M_\odot$ = 27\% $M_{\rm ej}$, 
N5def  : $0.16 M_\odot$ = 43\% $M_{\rm ej}$.
The energy released that may be converted to heat (including photons and electrons/positrons, excluding neutrinos which escape), remains modest compared to the kinetic energy of the ejecta (although somewhat less so for the models newly considered in this paper), as follows, 
N100ddt: $1.1\times10^{50}$~erg =  8\% $E_{\rm SN}$, 
N5ddt  : $1.8\times10^{50}$~erg = 12\% $E_{\rm SN}$, 
N100def: $6.6\times10^{49}$~erg = 11\% $E_{\rm SN}$, 
N5def  : $2.9\times10^{49}$~erg = 22\% $E_{\rm SN}$. 
Another~difference between the ddt and def models is the spatial distribution of the $^{56}{\rm Ni}$ synthesized. For ddt models it is concentrated in the core, with more asymmetries for N5ddt than for N100ddt. For def models it is spread more evenly over radii.

To check the impact of heating from radioactive decay on the SNR dynamics, we re-ran the SNR simulation under the assumption of local deposition of all the energy released, which constitutes an extreme case (see Section~4.2 and Appendix~C in Paper~I for the details). We had not seen a significant impact for N100ddt, for the other models also we observe only mild differences on the projected maps and power spectra, so that our main results above still hold. The effect of strong heating in the ejecta core is visible on the inner density structures, but not affecting much the dynamics of the wave fronts. 
Further investigating this would require diagnostics of the radioactive species still present deep within the un-shocked ejecta. 

Effects related to $^{56}{\rm Ni}$ will be the strongest in the first few years, and terminated within a hundred years (so well within the ejecta-dominated phase). Another radioactive element of interest if $^{44}{\rm Ti}$, the mass yields and energy deposits are very small, of the order of respectively $10^{-8} M_\odot$ and $10^{44}$~erg, but with a much longer life-time of about 60~yr. Detecting $^{44}{\rm Ti}$ in young Type~Ia SNRs has been a longstanding objective of hard X-ray / soft $\gamma$-ray missions, see \citet{Weinberger202044Remnants} and references therein for Tycho and Kepler SNRs.

\paragraph{Magnetic fields and particle acceleration.}

In Paper~I we already discussed the possible effects of the magnetic field and particle acceleration, which we do not expect to change the overall trends observed in this paper, when it comes to comparing different SN models with different ejecta morphology. That being said, we reckon that for a given SNR several physical processes may be at play at the same time at and around the shocks.
We have seen that, because of the initial asymmetry of the ejecta, the RS/FS ratio may vary systematically over the SNR, this is especially marked for the N5ddt model. The location of the shocks also depends on the compressibility of the plasma, which can be substantially affected by the presence of a non-thermal population of particles \citep[e.g.][]{Ferrand20103DAcceleration,Warren2013Three-dimensionalSNRs}. The presence of energetic particles increases the compressibility of the fluid, lowering the effective adiabatic index, which results in a narrower (and denser) shocked region. It~remains to be seen how the two effects will work together, and how our findings impact diagnostics of the efficiency of particle acceleration at SNR shock waves.

\paragraph{Impact of the ambient medium.}

We have observed that the N5 models lead to asymmetric SNRs, either with a dipolar structure for N5ddt, or with an offset structure for N5def. These effects in our simulations stem from the 3D distribution of the ejecta density. Similar effects, of systematic variations of the RS/FS ratio across directions and over time, could also be obtained from gradients in the density of the ambient medium \citep[e.g.][]{Orlando2007OnRemnants}. In this study we opted for the reference case of a uniform ISM, what we show is that fluctuations in the ISM are not necessary to obtain SNRs with non-canonical shapes during the ejecta-dominated phase. Older SNRs will most certainly have their shape affected by the ambient medium, which is not uniform on $>$pc scales. For recent simulation work on the impact of a turbulent or cloudy ISM on the development of the SNR, relevant to Type Ia SNe, see \citet{Wang2018TheRemnants,Peng2020SimulationsMedium} and \citet{Villagran2020EvolvingMedia}. 
\\~

\subsection{Comparison with observations}
\label{sec:discussion-obs}

There are only a few well-known Type Ia SNRs that are young (age $<1000$~yr) and nearby (spatially resolvable). In the Galaxy these are G1.9+0.3 ($\simeq 150$~yr), the youngest known \citep{Borkowski2013SupernovaG1.9+0.3}, Kepler = G4.5+6.8 = SN~1604 ($416$~yr), with a complicated morphology presumably from the progenitor's mass loss history \citep{Reynolds2007AInteraction,Burkey2013X-RaySupernova,Sato2017FreelyRemnant}, Tycho = G120.1+1.4 = SN~1572 ($448$~yr, \citealt{Warren2005Cosmic-RayObservations,Sato2017DirectRemnant}), and SN 1006 = G327.6+14.6 ($1014$~yr), with a characteristic bilateral morphology from the orientation of the magnetic field \citep{Uchida2013Asymmetric1006,Winkler2014ARemnant}. In the LMC we have SNR 0519-69.0, $450\pm200$~yr estimated from the shock dynamics \citep{Kosenko2010TheStudy}, $600\pm200$~yr from the light echo \citep{Rest2005LightCloud}, SNR 0509-67.5 \citep{Warren2004Raising050967.5}, $400\pm120$~yr from the light echo, and N103B, less than 850~yr from the measured expansion \citep{Williams2018TheN103B}, consistent with the light echo.

In Paper~I we discussed the case of Tycho's SNR with the N100ddt model, that can explain a standard SN~Ia. Maps in projection look more realistic in the 3Di versus 1Di setup. The SN model leads to power at large scales (small~$\ell$), which has been reported in observations \citep{Warren2005Cosmic-RayObservations}, but which had not been reproduced by simulations assuming smooth initial profiles. At age 500~yr, the simulated angular spectrum could be confused with an enhanced RT growth, when really it is the combination of both the decay from the SN modes and the growth from the RTI modes.

N5ddt is the most asymmetric model, from the SN phase into the early SNR phase. The resulting SNR at 500~yr has a clearly two-sided shell. Actually Tycho has a brighter side in X-rays too. 
Can initial asymmetries of the SN explain ejecta features observed in some young SNRs, like the iron-rich knots at the edge of Tycho \citep{Yamaguchi2017TheRemnant} and of Kepler \citep[][see in particular their Figure~8]{Sato2020ARemnant}? Such structures are on smaller scales than the dipole of N5ddt. From a purely hydrodynamic perspective, they would require ejecta bullets with a huge momentum to be found at the edge of the SNR (see \citealt{Tutone2020ThreeMixing} for a study of this aspect, in the context of core-collapse SNRs). In subsequent papers, we will have a closer look at the distribution of elements in space and in time.
Polarization studies \citep{Bulla2016PredictingSupernovae,Bulla2020WhiteInteraction} indicate that asymmetries are not as large that they would contradict the SN observations, which means that the discriminating power between the models is limited. This makes the striking differences in the SNR phase particularly relevant.

N5def may be a model for sub-energetic SNe events, which are less common than normal SNe~Ia. After incompleteness effects are taken into account \citet{Foley2013TypeExplosion} estimate that Type Iax SNe account for about 30\% of all SNe~Ia. We may or may not have observed a young and nearby remnant belonging to this class. \citet{Zhou2020ChemicalRemnant} just proposed, from a study of the elemental abundances, that SNR Sgr~A East = G0.0+0.0 is the remnant of a Type Iax SN, even though a core-collapse origin had been discussed in the past \citep{Maeda2002ACenter,Park2005AEast}. This appears to be the first such claim. However the age of this SNR remains uncertain, and it is located in the Galactic centre which is a complicated region. In any case it is interesting that there are specific predictions for the SNR phase for def models. 
One is the central over-density, although this part of the ejecta is not readily accessible to observations before it is shocked by the RS (which will take thousands of years). There actually is a class of SNRs with centrally peaked thermal emission in X-rays, the so-called mixed-morphology SNRs \citep{Rho1998Mixed-MorphologyRemnants}, also called thermal composites to distinguish them from plerionic composites where the central enhancement is due to a pulsar wind nebula (PWN). But these tend to be old SNRs, often interacting with molecular clouds, and other physical mechanisms are invoked like thermal conduction and evaporation. Besides, at present no SNR in the sample has been reliably confirmed to be of Type Ia origin. We however speculate that the SNR HB~9 = G160.9+02.6 may be one such remnant because of its high Fe/Si ratio \citep{Saito2020X-rayHB9}. 

Another notable feature of the def models is the presence of a network of seam lines on the ejecta surface, that are over-dense and so would be over-bright in X-rays. This appears to be an imprint of the deflagration dynamics, and so could possibly serve as an observational signature of this class of explosion models. In the ddt models, such structures are erased by the subsequent detonations. It~is currently not clear whether such structures exist in real SNRs.

To make more quantitative comparisons between our simulations and observations, we need to develop the appropriate analysis techniques, which is work in progress. One approach is to use techniques from harmonic analysis, similar to the spherical harmonics expansion used in Section~\ref{sec:results-healpix}, but in 2D slab geometry (for the SNR interior) or 1D circular geometry (for the outline of the SNR) as observed in projection. A~different approach is to use techniques from topology: \citet{Sato2019GenusSupernovae} applied the method of ``genus statistic" to characterize the clumpy structures in Tycho's SNR. The detailed comparison of observations with our models using such techniques may allow us to probe the explosion mechanisms for SNe~Ia.
\\~

\section{Conclusion}
\label{sec:conclusion}

This paper continues our study, started in Paper~I, of the imprint of the SN on the SNR, specifically from the point of view of 3D morphology (the SN-SNR connection has been done before, chiefly from the point of view of abundances). We have run numerical simulations of the young SNR phase, using as initial conditions four SN models of the thermonuclear explosion of a Chandrasekhar-mass WD: N100ddt, N5ddt, N100def, and N5def, taken from \citet{Seitenzahl2013Three-dimensionalSupernovae} and \citet{Fink2014Three-dimensionalSupernovae}. These models cover two explosion mechanisms: delayed detonation (ddt) and pure deflagration (def), and compare two setups for the central ignition: many ignition points (N100) or a few ignition points (N5). The properties of these models have been investigated in details for the SN phase \citep{Sim2013SyntheticSupernovae, Kromer20133DSupernovae, Bulla2016PredictingSupernovae, Bulla2020WhiteInteraction}, we present here for the first time how they evolve into the SNR phase. We simulated the hydro evolution for 500~yr, in an ambient medium assumed to be uniform in order to separate concerns. We analysed the morphology of the SNR over time, using 2D slices and maps in projection, as well as a spherical harmonics decomposition of the ejecta's edge (CD) and of the shocked region width (FS-RS).

We observed that N100 models produce different remnants than N5 models, and ddt models produce different remnants than def models. Compared to N100 models, N5 models have a strong dipole component, and produce asymmetric remnants. For N5def the entire shell, as bounded by the RS and FS, is off-centred, for N5ddt only the FS is off-centred and so the shocked region itself is asymmetric, with a thicker side, which creates a notably dipolar SNR as seen in projection. Pure deflagration models show the imprint of their specific mechanisms: the presence of a bound remnant, in the form of a central over-density, and large-scale plumes, in the form of seam lines on the ejecta surface. 
As expected these effects depend on time: the imprint of the SN is more visible early on, and by 500 years all models show a roughly spherical shell structure. As shown in Paper~I, by performing the angular decomposition of the radial variations of the CD, we can separate the SN modes (from the explosion), at large scales, that decay in time, and the SNR modes (from RTI), at small scales, that grow in time. By comparing the time evolution of the angular power between a setup with structured ejecta (3Di) and a setup with smooth ejecta (1Di), we are able to estimate the time until which the SN initial conditions dominate the shape of the SNR. It~ranges from 100 to 300~yr, and is significantly larger for N5 models than for N100 models.

This work shows the power of SNR simulations to discriminate between SN explosion models. To pursue this theoretical work, we will next try different kinds of models, like a double detonation of a sub-Chandrasekhar mass WD (the models by \citealt{Gronow2020SNeMechanism} and the D6 model by \citealt{Tanikawa2018Three-DimensionalCompanion}), or add a companion star \citep{Pakmor2008TheCompanions, Liu2012Three-dimensionalCompanions}. For a critical review of the many proposed scenarios for SNe~Ia, and how they relate with observations of SNRs, we refer the reader to \citet{Soker2019SupernovaeExplosions}. In subsequent papers we will present mock X-ray observations and analyze them in the same way as existing X-ray observations, starting with Tycho SNR. At that stage we will more fully exploit the 3D information from the SN models on the elemental abundances. 
\\~

\acknowledgments

This work is supported by JSPS Grants-in-Aid for Scientific Research ``KAKEN\-HI~A'' Grant Numbers JP19H00693. GF, SN, MO wish to acknowledge the support from the Program of Interdisciplinary Theoretical Mathematical Sciences (iTHEMS) at RIKEN, and the support from Pioneering Program of RIKEN for Evolution of Matter in the Universe (r-EMU).
IRS was supported by the Australian Research Council through Grant FT160100028.
The work of FKR is supported by the Klaus Tschira Foundation and by the Deutsche Forschungsgemeinschaft (DFG, German Research Foundation) -- Project-ID 138713538 -- SFB 881 (``The Milky Way System'', subproject A10).
TS was supported by the Japan Society for the Promotion of Science (JSPS) KAKENHI grant No. JP19K14739, the Special Postdoctoral Researchers Program, and FY 2019 Incentive Research Projects in RIKEN.
The authors thank the anonymous reviewer for comments that helped clarify the paper.

\facilities{Simulations were performed on the iTHEMS clusters at RIKEN.}

\software{HEALPix \citep{Gorski2005HEALPixSphere}, SciPy \citep{Virtanen2020SciPyPython}, Matplotlib \citep{Hunter2007Matplotlib:Environment}, VisIt \citep{Childs2012VisIt:Data}}
\\~


\appendix

\section{Evolution of N100def during the supernova phase}
\label{sec:N100def-SN}

\begin{figure}[p]
\centering
\includegraphics[width=\widthhealpix]{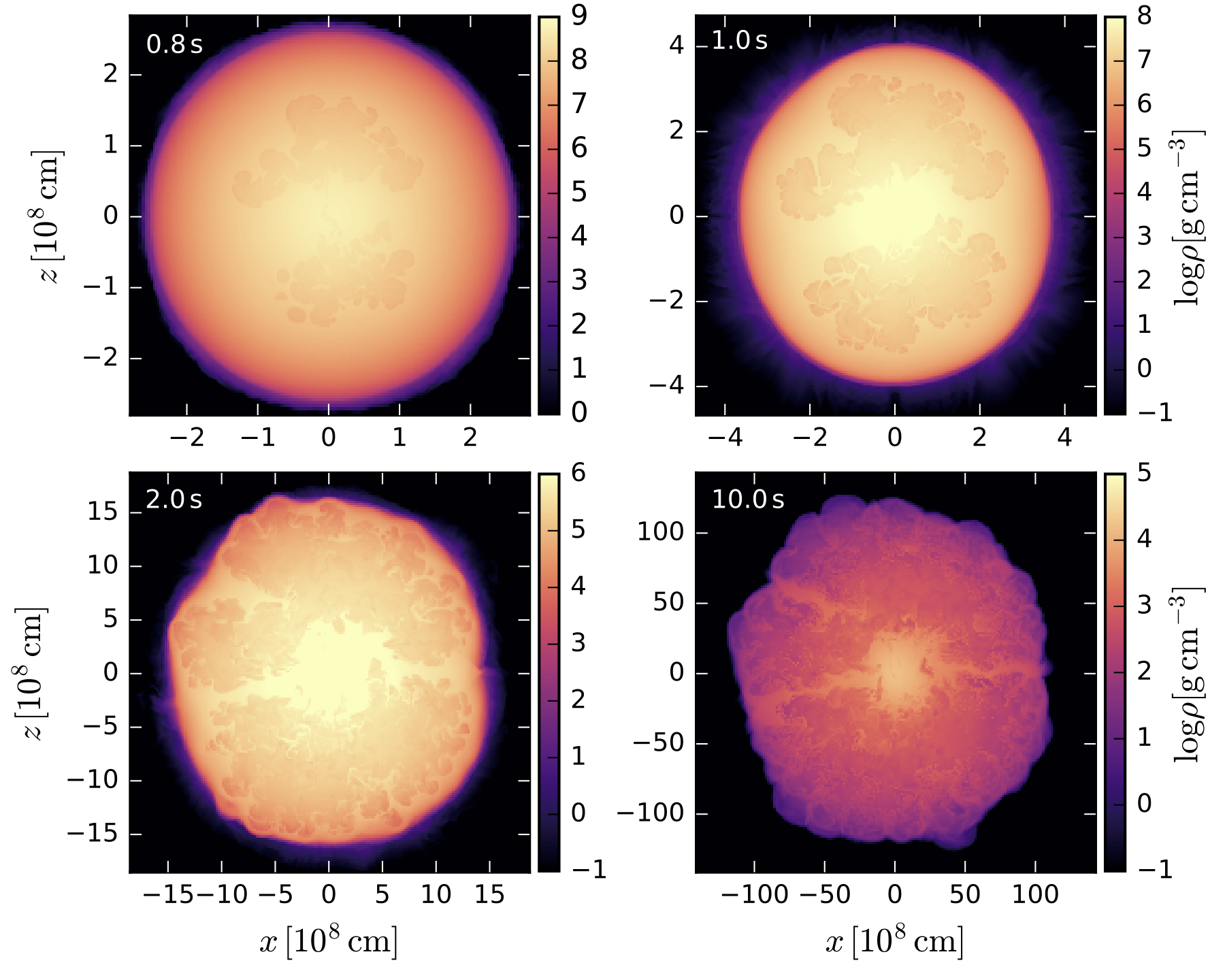}
\caption{Slices of the mass density for N100def during the SN phase, in the mid-plane along the y axis. Four times are shown: 0.8~s, 1.0~s, 2.0~s, and 10~s. An animation over time from 0 to 100~s is available (movie duration: 33~s).
\label{fig:N100def-SN-rho}}
\end{figure}

\begin{figure}[p]
\centering
\includegraphics[width=\widthhealpix]{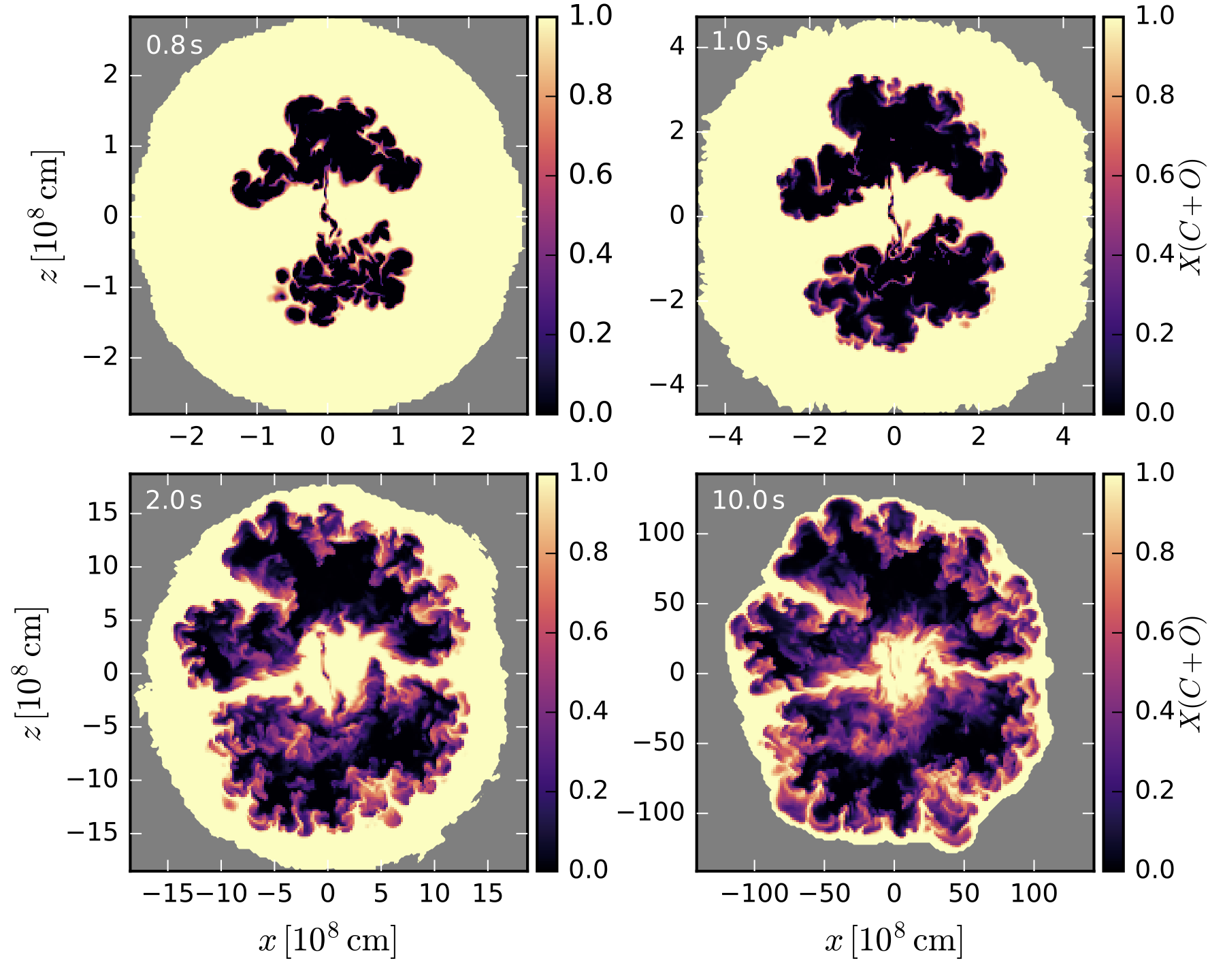}
\caption{Slices of the mass fraction of C+O for N100def during the SN phase, in the mid-plane along the y axis. Four times are shown: 0.8~s, 1.0~s, 2.0~s, and 10~s. An animation over time from 0 to 100~s is available (movie duration: 33~s). The gray area marks regions where the mass density falls below $0.1$ g.cm$^{-3}$.
\label{fig:N100def-SN-xCO}}
\end{figure}

In this section we show the temporal evolution of the SN simulation for model N100def, in order to elucidate the density structure used as initial conditions for the SNR simulation. Figure~\ref{fig:N100def-SN-rho} shows slices of the total mass density, while Figure~\ref{fig:N100def-SN-xCO} shows slices of the mass fraction of C+O. We see that large under-dense cells are growing over time, from the lack of C or O we can tell that these are combustion ashes. The over-dense filaments present at later times are boundaries between these plumes, still rich in C+O. We note that these plumes have irregular edges, which induces density variations as a function of direction inside the ejecta.

\newpage
\bibliography{references}{}
\bibliographystyle{aasjournal}

\end{document}